\documentclass[a4paper]{aa}
\usepackage{amsmath,amssymb,natbib,graphicx}
\graphicspath{{./}{./Figures/}}
\newcommand{\be}{\begin{equation}}
\newcommand{\ee}{\end{equation}}
\newcommand{\ellipticity}{\epsilon}
\newcommand{\size}{R}
\newcommand{\scale}{a}
\def\snr{{\mathcal{S}}}
\newcommand{\fluxtot}{F^{(0)}}
\newcommand{\param}{p}
\newcommand{\paramvector}{{\mathbf \param}}
\newcommand{\sigmasys}{\sigma_{\rm sys}}
\newcommand{\bs}{\boldsymbol}
\newcommand{\ellgal}{\ellipticity}
\newcommand{\snreff}{\snr_{\rm eff}}
\newcommand{\snrmax}{\snr_{\rm max}}
\newcommand{\snrmin}{\snr_{\rm min}}
\def\ssqobj#1{{\size^2_{\rm #1}}}

\def\variance#1{\left<\left|#1\right|^2\right>}

\begin{document}

\title{PSF calibration requirements\\
for dark energy from cosmic shear}

\titlerunning{PSF calibration requirements for dark energy from cosmic shear}
\authorrunning{Paulin-Henriksson et al.}

\author{S. Paulin-Henriksson\inst{1}, A. Amara\inst{1}, L. Voigt\inst{2}, A. Refregier\inst{1} and S.L. Bridle\inst{2}
}

\institute{
 Service d'Astrophysique, CEA Saclay, Batiment 709, 91191 Gif--sur--Yvette Cedex, France
\and
Department of Physics \& Astronomy, University College London, London, WC1E 6BT, U.K.
}

\date{Received 26 November 2007; accepted 4 April 2008}

\abstract
{% context
The control of systematic effects when measuring background galaxy shapes is one of the main challenges for cosmic shear analyses.
}
{% aims
Study the fundamental limitations on shear accuracy due to the measurement of the point spread function (PSF) from the finite number of stars that are available. We translate the accuracy required for cosmological parameter estimation to the minimum number of stars over which the PSF must be calibrated.
}
{% Methods
We characterise the error made in the shear arising from errors on the PSF. We consider different PSF models, from a simple elliptical gaussian to various shapelet parametrisations. First we derive our results analytically in the case of infinitely small pixels (i.e. infinitely high resolution), then image simulations are used to validate these results and investigate the effect of finite pixel size in the case of the elliptical gaussian PSF.
}
{% Results
Our results are expressed in terms of the minimum number of stars required to calibrate the PSF in order to ensure that systematic errors are smaller than statistical errors when estimating the cosmological parameters. On scales smaller than the area containing this minimum number of stars, there is not enough information to model the PSF. This means that these small scales should not be used to constrain cosmology unless the instrument and the observing strategy are optimised to make this variability extremely small. The minimum number of stars varies with the square of the star Signal-to-Noise Ratio, with the complexity of the PSF and with the pixel size. In the case of an elliptical gaussian PSF and in the absence of dithering, 2 pixels per PSF full width at half maximum (FWHM) implies a 20\% increase of the minimum number of stars compared to the ideal case of infinitely small pixels; 0.9 pixels per PSF FWHM implies a factor 100 increase.
}
{% Conclusions
In the case of a good resolution and a typical Signal-to-Noise Ratio distribution of stars, we find that current surveys need the PSF to be calibrated over a few stars, which may explain residual systematics on scales smaller than a few arcmins. Future all-sky cosmic shear surveys require the PSF to be calibrated over a region containing about 50 stars. Due to the simplicity of our models these results should be interpreted as optimistic and therefore provide a measure of a systematic `floor' intrinsic to shape measurements.
}

\keywords{Gravitational lensing - Cosmology: dark matter - Cosmology: cosmological parameters}

\maketitle

%%%%%%%%%%%%%%%%%%%%%%%%%%%%%%%%%%%%%%%%%%%%%%%%%%%
%%%%%%%%%%%%%%%%%%%%%%%%%%%%%%%%%%%%%%%%%%%%%%%%%%%
%%%%%%%%%%%%%%%%%%%%%%%%%%%%%%%%%%%%%%%%%%%%%%%%%%%

\section{Introduction}
\label{sec:intro}

Gravitational lensing by large scale structure (or `cosmic shear') has grown rapidly as a research field over the last decade  \citep{refregier03,hoekstra03,munshi06}. In the coming years, we can expect that the field will continue to make important contributions to cosmology. For instance, 
\cite{2006astro.ph..9591A} and \cite{2006Msngr.125...48P} have singled out cosmic
shear as potentially the most powerful probe for constraining dark
energy.

A great deal of work is currently under way to develop
techniques that will allow the maximum possible potential to be reached
and not limited by systematic measurement errors.  This can be done by designing and
building instruments with weak lensing as the central science driver.
This
approach leads to a top-down mission design process which begins with
science requirements, which are translated into technical requirements
and then into an instrument design.  This method for tackling the
problem of high accuracy weak lensing measurements is fundamentally
different to the bottom-up approach that is currently used. In the latter
approach an existing telescope is used and the science team
is assigned the task of extracting the maximum possible information from their surveys (until they hit the systematic limit of the
instrument).

The field of gravitational lensing lends 
itself most naturally to top-down design, since most (but not all) of
the systematic errors are not astronomical but are associated with the
instrument and the atmosphere. These can therefore be controlled
through instrument design and an optimised survey strategy. For this
reason, many of the ambitious future imaging surveys that are
currently under development have placed weak lensing as
primary
science driver, including the Dark UNiverse Explorer\footnote{http://www.dune-mission.net} (DUNE), 
the SuperNovae Acceleration Probe\footnote{http://snap.lbl.gov/} (SNAP),
the Panoramic Survey Telescope \& Rapid Response System\footnote{http://pan-starrs.ifa.hawaii.edu} (Pan-STARRS), 
the Dark Energy Survey\footnote{http://www.darkenergysurvey.org} (DES) 
and the Large Synoptic Survey Telescope\footnote{http://www.lsst.org} (LSST).
The issue is then to establish the instrumental
requirements needed to reach the full statistical potential of the
survey.

In \cite{2007MNRAS.381.1018A}, a wide range of survey parameters, such as area and depth, were considered.  Their impact on the statistical potential of a lensing survey was calculated and summarised in a scaling relation that can be used for survey designs to trade-off one property of a survey against another.  In brief, this work finds that, once the median redshift of a survey is sufficiently high ($z \gtrsim 0.7$), the optimal survey strategy is to make the lensing survey as wide as possible. Similar results have been found by \cite{2006MNRAS.373..105H}.
In the same spirit, \cite{2007arXiv0710.5171A} looked at the requirements an ultra-wide field survey
places on the control of systematics and concluded with a set of scaling relations that show the tolerance on residual systematic errors as a function of survey parameters. In particular, two types of shape measurement systematics were considered: multiplicative and additive systematics which are, respectively, correlated and uncorrelated with the lensing signal.
\cite{2006MNRAS.366..101H} have also studied the impact of multiplicative and additive errors.

In the present study, we link the above systematic requirements to errors associated with measurement of the point spread function (PSF) of the instrument. Indeed, since galaxy images need to be PSF-corrected before shapes can be measured, errors in the estimation of the PSF are propagated into errors on galaxy shapes and can mimic the shear signal. The PSF calibration is driven by 3 factors:
\begin{enumerate}
\item \textbf{The PSF model:} The model that is chosen to describe the PSF will never perfectly describe the true PSF hence the choice of PSF model introduces an `a priori' systematic;

\item \textbf{The interpolation scheme:}
The PSF will vary over the field which means that the PSF at a galaxy position needs to be interpolated from the PSFs of the nearby stars;

\item \textbf{The finite information available from each star:} A star provides an image of the PSF that is noisy and pixelated. Thus, the PSF information that we are able to extract from each star is finite. As a result, to reach high precision, it is necessary to combine the information from several stars.
\end{enumerate}
This paper is devoted to the last factor. We quantify the accuracy of the PSF calibration with analytical predictions in the case of infinitely small pixels and use image simulations to quantify the pixelation effects. Previous work has also studied the impact that PSF errors are likely to induce in cosmic shear measurements. For instance,  \cite{2007arXiv0710.3355S} looked specifically at the PSF from the current  SNAP design and translated this into systematic errors.  The study that we present here sets out to be more general  and uses simplified PSFs to try and quantify the systematic floor of an instrument. 
We investigate the extent to which the pixel scale will degrade the amount of information relative to our analytic predictions but we do not explore the optimal methods for combining multiple exposures, such as the one proposed by \cite{2006JCAP...02..001J}. \cite{2007astro.ph..3471H} have studied the impact of the pixel scale on the number of useful galaxies (assuming the PSF is known perfectly). While they find a weak dependence, 
we expect that 
the driving factor for pixel scale is shape measurement systematics rather than statistical errors.

For a survey with a given statistical potential
we use the results of \cite{2007arXiv0710.5171A} to find the upper limit on the systematic errors in the shears such that the induced bias in cosmological parameters remain subdominant compared to the statistical (marginalised Fisher matrix) errors 
when estimating cosmological parameters. We then convert the requirements on shear systematics into the minimum number of stars ($N_*$) needed to measure the PSF to the level of accuracy required.
On scales smaller than the area containing $N_*$, there is not enough information from the stars to calibrate the PSF. Therefore, the systematic errors in the cosmological parameters can be dominant over the statistical ones. This means that these small scales should not be used to constrain cosmology unless the PSF variability is known to be extremely small. 
$N_*$ clearly depends on the Signal-to-Noise Ratio distribution of stars, since bright ones contain more information about the PSF than faint noisy ones. 
It also depends on the stability and the complexity of the PSF, 
characterised by the number of degrees of freedom that must be estimated or interpolated from the stars. Each degree of freedom increases $N_*$.

\begin{figure}
\resizebox{\hsize}{!}{
\includegraphics{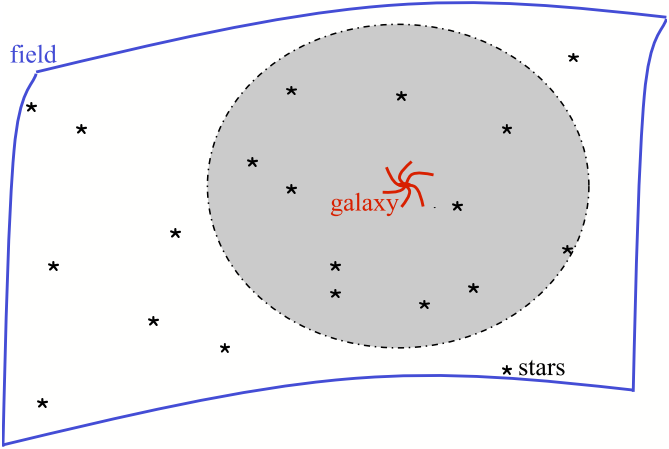}
}
\caption{
\label{fig:cartoonns}
Illustration of the required number of stars $N_*$ characterised in this paper.
When measuring the shape of a galaxy (marked by a red spiral),
the PSF needs to be calibrated with at least $N_*$ nearby stars (black asterisks) contained in
the shaded region. In this example $N_*=11$. 
On scales smaller than this there is not enough information coming from the stars to measure the PSF variations.
}
\end{figure}

Figure \ref{fig:cartoonns} shows a graphical example to illustrate the calibration requirement we are considering. We see a central galaxy surrounded by several stars. 
The grey shaded region shows the area over which the PSF must be 
calibrated 
and in this example contains 11 stars. 
Stars do not provide enough information to model the PSF variations on scales smaller than this area.

This paper is organised as follows. Section \ref{sec:systematics} explains the weak lensing and cosmological context and summarises the issue of accurate PSF calibration. Section \ref{sec:analyticalmodel} shows the analytical predictions of the PSF calibration accuracy, expected in the case of infinitely small pixels. Section \ref{sec:simulations} describes the simulations we use to validate these predictions and extend them to finite pixels. In section \ref{sec:discussion} we give the final accuracy of the PSF calibration and derive the number of stars it requires.

%%%%%%%%%%%%%%%%%%%%%%%%%%%%%%%%%%%%%%%%%%%%%%%
%%%%%%%%%%%%%%%%%%%%%%%%%%%%%%%%%%%%%%%%%%%%%%%
%%%%%%%%%%%%%%%%%%%%%%%%%%%%%%%%%%%%%%%%%%%%%%%

\section{Weak lensing}
\label{sec:systematics}

\subsection{Shear Measurement}
\label{sec:shearmeasurement}

Weak gravitational shear is locally estimated using the shapes of background galaxies \citep[for a review, see e.g.][]{2001PhR...340..291B}.
Shear estimation methods can be divided into two families: those computing an estimator of the shear from a set of weighted sums over pixel values, for instance the common `KSB+' method \citep{1995ApJ...449..460K,1997ApJ...475...20L,1998ApJ...504..636H},
and those fitting a model to the observed galaxy shape and deriving an estimator from the model. This includes the methods proposed by \cite{1999A&A...352..355K,2002AJ....123..583B,2002sgdh.conf...29R} and \cite{voigtb08}.

In this paper, we need a general formalism to propagate the error on the PSF into an error on the shear estimate. For this purpose, 
we consider as shape parameters the 2 component ellipticity $\bs{\ellipticity}$ and the squared radius 
$\size^2$, both defined using the
unweighted second order moments of the galaxy. 
For an object with surface brightness $f(x_1,x_2)$, the total flux $\fluxtot$ is the zeroth order moment of the surface brightness:
\be
\label{eq:fluxdef}
\fluxtot=\int\,d^2\bs{x}\,f(\bs{x})\,,
\ee
the centroid $\bs{x}^{\rm cen}$ is given by the first order moments divided by the total flux:
\be
\label{eq:centroiddef}
x^{\rm cen}_i = \frac{F_i^{(1)}}{\fluxtot} = \frac{1}{\fluxtot} \int\,d^2\bs{x}\,x_i\,f(\bs{x})
\ee
and the quadrupole moment matrix is given by the second order moments divided by the total flux:
\be
\label{eq:quadrupoledef}
Q_{ij}=\frac{F_{ij}^{(2)}}{\fluxtot}=\frac{1}{\fluxtot} \int\,d^2\bs{x}\,(x_i-x_i^{\rm cen})(x_j -x_j^{\rm cen})\,f(\bs{x})\,.
\ee
The square rms radius $\size^2$ and the 2 component ellipticity $\bs{\ellipticity}=[\ellipticity_1,\ellipticity_2]$ are defined by:
\begin{eqnarray}
\label{eq:sizedef}
\size^2 & \equiv & Q_{11} + Q_{22}\;,\\
\label{eq:ellipticitydef1}
\ellipticity_1 & \equiv & \frac{Q_{11}-Q_{22}}{Q_{11} + Q_{22}}\;,\;\ellipticity_2 \equiv \frac{2Q_{12}}{Q_{11} + Q_{22}}\;,\\
\label{eq:ellipticitydef2}
\ellipticity & \equiv & \sqrt{\ellipticity^2_1+\ellipticity^2_2}\;.
\end{eqnarray}
Consider now we have unbiased estimators of $\size^2$ and $\bs{\ellipticity}$ with variations $\delta\size^2$ and $\bs{\delta\ellipticity}$ around the true values. We adopt the following definitions:
\begin{eqnarray}
\label{eq:sigmasizedef}
\variance{ \delta\size^2 } & \equiv & \sigma^2[\size^2]\;,\\
\label{eq:sigmaelldef}
\variance{\bs{\delta\ellipticity}} & \equiv & 2\sigma^2[\ellipticity]\;.
\end{eqnarray}
The factor 2 in equation \ref{eq:sigmaelldef} is due to the fact that $\bs{\ellipticity}$ has 2 components and we define $\sigma[\ellipticity_{\rm PSF}]$ as the standard deviation of one of the components.

The weak gravitational shear $\bs{\gamma}=[\gamma_1,\gamma_2]$ can then be shown to be estimated using:
\begin{eqnarray}
\label{eq:gammadef}
\widehat{\bs{\gamma}} & = & (P^\gamma)^{-1}\,\bs{\ellipticity}_{\rm gal}\;,
\end{eqnarray}
where
\begin{eqnarray}
\label{eq:pgammadef}
P^\gamma & = & 2-<\left|\bs{\ellipticity}_{\rm gal}\right|^2> \approx 1.84\;.
\end{eqnarray}
The symbol `$\widehat{\quad}$' indicates an estimator so that \mbox{$<\widehat{\bs{\gamma}}>=\bs{\gamma}$}, $P_\gamma$ is the shear susceptibility and the subscript `gal' corresponds to values measured on a galaxy. The value of \mbox{$\sqrt{<\left|\bs{\ellipticity}_{\rm gal}\right|^2>}\approx 0.4$} comes from the typical ellipticity distribution in current data sets.

In practice, the measurement of the galaxy ellipticity is uncertain because of the noise in the image. 
We write the induced error as 
$\bs{\delta\ellgal}^{\rm noise}$, which has a null average \mbox{$<\bs{\delta\ellgal}^{\rm noise}>=0$}.
There may also be a systematic effect $\bs{\delta\ellgal}^{\rm sys}$
so that the estimated ellipticity is
\begin{eqnarray}
\label{eq:ellipticitydef3}
\widehat{\bs{\ellipticity}}_{\rm gal} & = & \bs{\ellipticity}_{\rm gal} + \bs{\delta\ellgal}^{\rm sys} + \bs{\delta\ellgal}^{\rm noise}\;.
\end{eqnarray}
In practice
$\bs{\delta\ellgal}^{\rm sys}$ may be due to several factors. 
For instance, the STEP collaboration \citep{2006MNRAS.368.1323H,2007MNRAS.376...13M,spacestep} investigate the contribution to $\bs{\delta\ellgal}^{\rm sys}$ from imperfections in the shear measurement methods. In a different spirit, this paper focusses on the contribution to $\bs{\delta\ellgal}^{\rm sys}$ from the limited information available on the PSF due to the photon noise in the star images. We identify the variance $\variance{ \bs{\delta\ellgal}^{\rm sys} }$ with the quantity $\sigma_{\rm sys}^2$ examined in \cite{2007arXiv0710.5171A}, which indicates the variance of the systematic errors
\be
\label{eq:sigmasysdef1}
\sigma_{\rm sys}^2 \equiv \left(P^\gamma\right)^{-2} \variance{ \bs{\delta\ellipticity}^{\rm sys} }\;.
\ee
We translate an upper limit on $\sigmasys^2$ into constraints on the number stars that are needed to calibrate the PSF.

%%%%%%%%%%%%%%%%%%%%%%%%%%%%%%%%%%%%%%%%%%%%%%%
%%%%%%%%%%%%%%%%%%%%%%%%%%%%%%%%%%%%%%%%%%%%%%%

\subsection{Systematics from PSF calibration}
\label{sec:sysfrompsfcalib}
Since a galaxy image needs to be PSF corrected before its shape can be measured, errors in the estimation of the PSF propagate into an error $\bs{\delta\ellipticity}^{\rm sys}$ in the measured ellipticity. Here, we consider errors in the PSF radius parameter $\size_{\rm PSF}$ and in the 2 component PSF ellipticity $\bs{\ellipticity}_{\rm PSF}$. To first order and for the unweighted moments (definitions given by equations \ref{eq:quadrupoledef} to \ref{eq:ellipticitydef2}), we have (see Appendix \ref{sec:propagation} for details):
\be
\label{eq:errorpropagationpsfintogal}
\bs{\delta\ellipticity}^{\rm sys} \simeq 
\left( \bs{\ellipticity}_{\rm gal} - \bs{\ellipticity}_{\rm PSF} \right)
 \frac{\delta\left(\ssqobj{PSF}\right)}{\ssqobj{gal}} - \left(\frac{\size_{\rm PSF}}{\size_{\rm gal}}\right)^2 \bs{\delta\ellipticity}_{\rm PSF}
\ee
where $\size_{\rm gal}$ and $\size_{\rm PSF}$ are the radius parameters of the galaxy and the PSF respectively. The first two terms show that systematics due to an error on the PSF size are proportional to the ellipticities of the galaxy and of the PSF. In the following, we show that the latter should be optimised to be as small as possible. 
The last term shows that systematics due to an error on the PSF ellipticity are proportional to the squared ratio between the PSF and galaxy sizes.

Combining equations \ref{eq:sigmasysdef1} and \ref{eq:errorpropagationpsfintogal} gives the propagation of PSF errors into $\sigma_{\rm sys}^2$. For this purpose we make the following simplifications:
\begin{enumerate}
\item The galaxy is not correlated with the PSF (i.e. the crossed-terms \mbox{$<\bs{\ellipticity}_{\rm gal}.\bs{\ellipticity}_{\rm PSF}>$} and \mbox{$<\bs{\ellipticity}_{\rm gal}.\bs{\delta\ellipticity}_{\rm PSF}>$} are equal to 0).
\item The error on the PSF ellipticity ($\bs{\delta\ellipticity}_{\rm PSF}$) and the PSF ellipticity itself ($\bs{\ellipticity}_{\rm PSF}$) are not correlated. This is warranted by the fact that, in the assumed small ellipticity regime, $\bs{\delta\ellipticity}_{\rm PSF}$ does not have any preferred direction, implying \mbox{$<\bs{\ellipticity}_{\rm PSF}.\bs{\delta\ellipticity}_{\rm PSF}>=0$}.

\item We assume that the ellipticity and the inverse squared radius of the galaxy are also uncorrelated. More exactly we assume:
\be
\label{eq:assumption}
\variance{ \frac{\bs{\ellipticity}_{\rm gal}}{\size^2_{\rm gal}} } \simeq \variance{ \bs{\ellipticity}_{\rm gal} } \variance{ 1/\size^2_{\rm gal} }\;.
\ee
This assumption is reasonable for our work on the PSF calibration presented in this paper.
\end{enumerate}
With these simplifications, we obtain:
\begin{eqnarray}
\label{eq:sigmasysdef}
\sigma_{\rm sys}^2 & = & \left(P^\gamma\right)^{-2} \left<\left(\frac{\size_{\rm PSF}}{\size_{\rm gal}}\right)^4\right> \times
     \Bigg[2\sigma^2[\ellipticity_{\rm PSF}]\nonumber\\
 & & +\left( \variance{ \bs{\ellipticity}_{\rm gal} } + \variance{ \bs{\ellipticity}_{\rm PSF} } \right)
     \left(\frac{\sigma[\size^2_{\rm PSF}]}{\size^2_{\rm PSF}}\right)^2
     \Bigg]\;.
\end{eqnarray}
This equation confirms the intuition that the PSF ellipticity of a cosmic shear survey should be small. We see that the last ellipticity term inside the brackets is 
$\variance{\bs{\ellipticity}_{\rm PSF}}$
and should be reduced when optimising the survey to have $\ellipticity_{\rm PSF}$ as small as possible.

For instance, to reach the requirements of \mbox{$\sigmasys^2 \lesssim 10^{-7}$}, with a typical well sampled cosmic shear survey with $\variance{\bs{\ellipticity}_{\rm gal}}=0.16$, $R_{\rm gal} \ge 1.5\,R_{\rm PSF}$ and 
\mbox{$\ellipticity_{\rm PSF}\lesssim 0.05$}, would require:
\begin{eqnarray}
\frac{\sigma[\size^2_{\rm PSF}]}{\size^2_{\rm PSF}} 
& \lesssim & 10^{-3}\;,\\
\sigma[\ellipticity_{\rm PSF}]
& \lesssim & 10^{-3}.
\end{eqnarray}
In the following, we discuss how these upper limits translate into requirements on the PSF calibration.

%%%%%%%%%%%%%%%%%%%%%%%%%%%%%%%%%%%%%%%%%%%%%%
%%%%%%%%%%%%%%%%%%%%%%%%%%%%%%%%%%%%%%%%%%%%%%%%
%%%%%%%%%%%%%%%%%%%%%%%%%%%%%%%%%%%%%%%%%%%%%%%%

\section{Analytical model}
\label{sec:analyticalmodel}

In this section, we consider the general problem of fitting a 2D model of the PSF to a pixelated image of a star with additive uncorrelated gaussian noise using $\chi^2$ minimisation. This allows us to derive a number of analytic results in the limit of infinitely small pixels (i.e. infinitely high resolution), which will serve as a useful comparison base for the numerical simulation studies
presented in section \ref{sec:simulations}.

%%%%%%%%%%%%%%%%%%%%%%%%%%%%%%%%%%%%%%%
%%%%%%%%%%%%%%%%%%%%%%%%%%%%%%%%%%%%%%%

\subsection{General 2D fit}
\label{sec:general2dfit}
Consider the PSF surface brightness as a function of position on the image $\bs{x}$ to be described by a model $m(\bs{x};\paramvector)$ parameterised by parameters $\paramvector$. The observed surface brightness of a star is \mbox{$f(\bs{x})=m(\bs{x};\paramvector({\rm PSF}))+n(\bs{x})$} where $\paramvector({\rm PSF})$ is the true values of the parameters, $n(\bs{x})$ is the noise, which is assumed to be uncorrelated (from pixel to pixel) and gaussian with \mbox{$<n(\bs{x})>=0$} and \mbox{$<n(\bs{x})^2>=\sigma_n^2$} 
is assumed constant across the image. 
The usual \mbox{$\chi^2$-functional} is given by
\be
\label{eq:chisqdef}
\chi^2(\paramvector)= \sum_k \sigma_{n}^{-2} \left[ f(x_k) - m(x_k;\paramvector) \right]^{2},
\ee
the sum being over all pixels $k$ in the image. The usual estimator $\hat{\paramvector}$ is constructed by requiring that \mbox{$d\chi^2/d\paramvector=0$} when evaluated at $\hat{\paramvector}$. The covariance matrix of the estimated parameters is given by the inverse of the Fisher matrix:
\be
{\rm cov}[\param_i,\param_j] \simeq (\mathcal{F}^{-1})_{ij}
\ee
with
\be
\label{eq:fishermatrix}
\mathcal{F}_{ij} = \sigma_n^{-2} \sum_k \frac{\partial
m(x_k;\paramvector)}{\partial \param_i}\frac{\partial m(x_k;\paramvector)}{\partial \param_j}
\ee
and the variance of a function $P(\paramvector)$ of the fitted parameters is:
\be
\label{eq:sigmap}
\sigma^2[P] \simeq \sum_{i,j}\frac{\partial P}{\partial \param_i}\frac{\partial P}{\partial \param_j}{\rm cov}[p_i,p_j]\;.
\ee
In the following, we consider 2 such functions of the parameters $\boldsymbol{p}$: the rms radius squared $\size^2$ and the 2 component ellipticity $\bs{\ellipticity}$ (as defined in equations \ref{eq:sizedef} to \ref{eq:ellipticitydef2}), in the case of a simple elliptical gaussian PSF (section \ref{sec:ellgauss}) and in the case of more complex PSFs defined with shapelets (section \ref{sec:shapelets}). We define the associated dimensionless complexity factors $\psi_{\size^2}$ and $\psi_{\ellipticity}$ such as:
\begin{eqnarray}
\label{eq:sigsize2}
\sigma[\size^2] & \equiv & \frac{\size^2}{\snr}\,\psi_{\size^2}\;,\\
\label{eq:sigell2}
\sigma[\ellipticity] & \equiv & \frac{1}{\snr}\,\psi_{\ellipticity}
\end{eqnarray}
where $\snr$ is the Signal-to-Noise Ratio defined as:
\be
\label{eq:snrdef}
\snr = \frac{\fluxtot}{\sigma[\fluxtot]}\;.
\ee
$\sigma[\fluxtot]$ is the standard deviation of the total flux. 
Basically, $\psi_{\size^2}$ and $\psi_{\ellipticity}$ characterise the numbers of degrees of freedom in the PSF model associated with $\size^2$ and $\bs{\ellipticity}$.

In the limit of infinitely small pixels, the sum over pixels in equations \ref{eq:chisqdef} and \ref{eq:fishermatrix} can be replaced by a continuous integral over the object. A number of analytical results can be derived. In section \ref{sec:ellgauss} we derive these for an elliptical gaussian PSF and  in section \ref{sec:shapelets} we study more complex PSFs described with shapelets.

%%%%%%%%%%%%%%%%%%%%%%%%%%%%%%%%%%%%%%%%%%%%%%
%%%%%%%%%%%%%%%%%%%%%%%%%%%%%%%%%%%%%%%%%%%%%%

\subsection{Elliptical gaussian}
\label{sec:ellgauss}
We first consider a 2D elliptical gaussian model parameterised as:
\begin{eqnarray}
m(\bs{x};\bs{p})\!\! & = &\!\! \frac{A}{2\pi \sqrt{\scale_1\,\scale_2}} \exp \left[ -\frac{1}{2} ( \bs{x}-\bs{x}^a)^{T} {\mathbf Q}^{-1} (\bs{x}-\bs{x}^a) \right],\\
\bs{p} & = & (\bs{x}^a,A,a_1,a_2,\alpha)
\end{eqnarray}
where $\scale_1$ and $\scale_2$ are the rms major and minor axes of the
gaussian, respectively, $A$ is a parameter which controls the
amplitude, $\bs{x}^a$ is the (true) centroid, and $^T$ stands for the
transpose operator. The total flux (as defined in equation \ref{eq:fluxdef}) is:
\be
\label{eq:fluxgauss}
\fluxtot=A\sqrt{\scale_1\,\scale_2}\;,
\ee
the centroid (as defined in equation \ref{eq:centroiddef}) is:
\be
x_i^{\rm cen}=x_i^a
\ee
and the quadrupole moment (as defined in equation \ref{eq:quadrupoledef}) is:
\be
{\mathbf Q}={\mathbf R}(\alpha)^{T}
\left( \begin{array}{cc}  \scale_1^2 & 0 \\ 0 & \scale_2^2 \end{array}  \right)
{\mathbf R}(\alpha)
\ee
where $\alpha$ is the position angle of the major axis
counter-clockwise from the $x$-axis and:
\be
{\mathbf R}(\alpha)= \left(
\begin{array}{cc} \cos \alpha & \sin \alpha \\ -\sin \alpha & \cos \alpha \end{array}
\right)
\ee
is the rotation matrix which aligns the coordinate system with the major
axis. 
The ellipticity and the squared radius (as defined in equations \ref{eq:sizedef} to \ref{eq:ellipticitydef2}) are:
\begin{eqnarray}
\label{eq:sizegauss}
\size^2 & = & \scale_1^2+\scale_2^2\;,\\
\label{eq:ellipticitygauss}
\ellipticity & = & \frac{\scale_1^2-\scale_2^2}{\size^2}\;.
\end{eqnarray}

This parametrisation is particularly convenient because for infinitely small pixels the Fisher matrix of the parameters, given by equation \ref{eq:fishermatrix}, is diagonal with diagonal elements:
\be
F_{ii} = \snr^2
\left( \frac{1}{\scale_1^2},  \frac{1}{\scale_2^2}, \frac{2}{A^2}, \frac{1}{\scale_1^2},\frac{1}{\scale_2^2},
\frac{(\scale_1^2-\scale_2^2)^2}{2 \scale_1^2 \scale_2^2} \right)\;.
\end{equation}
Consequently, with equation \ref{eq:sigmap}, the errors on the major and minor axes are:
\begin{eqnarray}
\label{eq:sigmascalegauss}
\sigma[\scale_i] & = & \frac{a_i}{\snr}.
\end{eqnarray}
From equations \ref{eq:sizegauss}, \ref{eq:ellipticitygauss} and \ref{eq:sigmascalegauss}, it follows that:
\begin{eqnarray}
\label{eq:analyticalgaussiansigmasize}
\sigma[\size^2] & = & \frac{2}{\snr}\sqrt{\scale_1^4+\scale_2^4}\;,\\
\label{eq:analyticalgaussiansigmaell}
\sigma[\ellipticity] & = & \frac{\scale_1^2\,\scale_2^2}{\size^4}\frac{4\sqrt{2}}{\snr}\;.
\end{eqnarray}
We generalise this simple elliptical gaussian model to more complex PSFs in the following section and 
test these equations for finite pixels using simulations in section \ref{sec:simulations}.

Under the approximation of small ellipticity ($\ellipticity \lesssim 0.1$), the dimensionless complexity factors $\psi_{\size^2}$ and $\psi_{\ellipticity}$ (defined by equations \ref{eq:sigsize2} and \ref{eq:sigell2}) are constant (i.e. do not depend on the object). Indeed, $\ellipticity\lesssim 0.1$ implies $a_1\simeq a_2$ and therefore in equations \ref{eq:analyticalgaussiansigmasize} and \ref{eq:analyticalgaussiansigmaell}, we have $\sqrt{a_1^4+a_2^4}\simeq\size^2/\sqrt{2}$ and $4a_1^2a_2^2\simeq \size^4$. This leads to:
\be
\psi_{\size^2}({\rm gauss}) = \psi_{\ellipticity}({\rm gauss}) = \sqrt{2}\;.
\ee

%%%%%%%%%%%%%%%%%%%%%%%%%%%%%%%%%%%%%%%%%%%%%%%%%%%%%%%%%%
%%%%%%%%%%%%%%%%%%%%%%%%%%%%%%%%%%%%%%%%%%%%%%%%%%%%%%%%%%

\subsection{Shapelet model}
\label{sec:shapelets}

To explore a wide variety of possible PSFs we consider a basis set that allows for complexity.
In this framework, the shapelet basis 
sets are 
particularly convenient because: (i) shapelets 
provide some orthonormal basis sets that have 
already been studied in number of publications \citep{2003MNRAS.338...35R,2003MNRAS.338...48R,2005MNRAS.363..197M}, tested on the STEP2 and STEP3 simulated data \citep{2007MNRAS.376...13M,spacestep} and used for the weak lensing analysis of the Canada-France-Hawai-Telescope-Legacy-Survey (CFHTLS) data set in the framework of a comparison with X-ray surveys \citep{2007arXiv0712.3293B}; (ii) as we show in the following, in the case of simple objects the standard deviations of the squared radius $\sigma[\size^2]$ and ellipticity $\sigma[\ellipticity]$ can be written as proportional to some complexity factors containing all the information about the basis.

A shapelet basis set is characterised by 2 parameters: $n_{\rm max}$, the maximum order of the functions in the basis and the scale parameter $\beta$. The surface brightness $f(\bs{x})$ of an object is described by:
\be
\label{eq:shap}
f(\bs{x})=\sum_{n=0}^{n_{\rm max}}\sum_{m=-n}^{n}f_{n,m}\,\chi_{n,m}(\bs{x},\beta)
\ee
with $\chi_{n,m}$ the polar shapelet functions 
and $f_{n,m}$ the (complex) polar shapelet coefficients of the object. 
The scale of the oscillations described by the function $\chi_{n,m}$ is proportional to $1/(n+\left|m\right|)$. It is often convenient to impose a lower limit to the scales which are described. This corresponds to setting the coefficients with 
$\left|m\right|>n_{\rm max} - n$ to 0. This configuration 
is called `diamond' and is only defined for an even $n_{\rm max}$. 
For simple objects with reasonable substructures and tails, 
the number of coefficients required is small when $\size^2 \sim 2\beta^2$. 
This relation is exact for a circular gaussian represented with $n_{{\rm max}}=0$.

We show in Appendix \ref{sec:appendixshapelets} that, under the approximation of small ellipticity ($\ellipticity \lesssim 0.1$) and for stars described with a judicious scale parameter where $\size^2=2 \beta^2$, the dimensionless complexity factors defined by equations \ref{eq:sigsize2} and \ref{eq:sigell2} (\mbox{$\sigma[\size^2]\equiv\size^2\psi_{\size^2}/\snr$} and \mbox{$\sigma[\ellipticity]\equiv\psi_{\ellipticity}/\snr$}) depend only on the basis (i.e. not on the object itself) and are given by:
\begin{eqnarray}
\label{eq:psidef1}
\psi_{\size^2} & \simeq & \sqrt{\frac{N}{3}(N+1)}\;,\\
\label{eq:psidef2}
\psi_{\ellipticity} & \simeq & \sqrt{\frac{N}{3}(N+4)}\quad{\rm without}\;{\rm diamond}\;,\\
\label{eq:psidef3}
\psi_{\ellipticity} & \simeq & \sqrt{\frac{N}{3}(N-2)}\quad{\rm with}\;{\rm diamond}
\end{eqnarray}
where $N$ is the largest even integer lower than or equal to $n_{\rm max}$. 
Figure \ref{fig:stddev.vs.nmax} illustrates equations \ref{eq:sigsize2}, \ref{eq:sigell2} and \ref{eq:psidef1} to \ref{eq:psidef3} by showing $\sigma[\ellipticity]$ and $\sigma[\size^2]$ as a function of $n_{\rm max}$.
$\psi_{\ellipticity}$ and $\psi_{\size^2}$ are given in table \ref{tab:psf} for 16 bases with $n_{\rm max}$ between 4 and 16 and for the elliptical gaussian model presented in the previous section.

%%%%%%%%%%%%%%%%%%%%%%%%%%%%%%%%%%%%%%
%%%%%%%%%%%%%%%%%%%%%%%%%%%%%%%%%%%%%%
%%%%%%%%%%%%%%%%%%%%%%%%%%%%%%%%%%%%%%

\section{Simulations}
\label{sec:simulations}

In this section, we use image simulations to test and validate the analytic predictions of the previous section. These simulations 
allow us to investigate the effects of pixelation 
in the case of an elliptical gaussian PSF. 
In this simple case
we show that pixelation degrades the accuracy of the shape measurements and thus, if the pixel scale is increased, the number of stars needed to calibrate the PSF also increases.

Star images are simulated for: (i) a simple elliptical gaussian PSF and (ii) complex PSFs derived from the STEP III simulations \citep{spacestep}. In all cases the pixelated images are produced by distributing the star centroids randomly and uniformly over the central pixel. Additive gaussian noise is then added to each image and the model fitting is done using $\chi^2$ minimisation (for the work presented here, we have not added Poisson noise, which arises from the object itself). gaussian stars are simulated at high resolution (each pixel is considered as the sum of $7\times 7$ sub-pixels), while shapelet stars are simulated using the analytic pixel integration scheme proposed in the online IDL pipeline\footnote{http://www.astro.caltech.edu/$\widetilde{\;}$rjm/shapelets} by \cite{2005MNRAS.363..197M}.

%%%%%%%%%%%%%%%%%%%%%%%%%%%%%%%%%%%%%%%%%%%%%%%%%
%%%%%%%%%%%%%%%%%%%%%%%%%%%%%%%%%%%%%%%%%%%%%%%%%

\subsection{Elliptical gaussian PSF}
\label{sec:simellgauss}
\begin{figure*}
\begin{center}
\includegraphics[height=1.5cm]{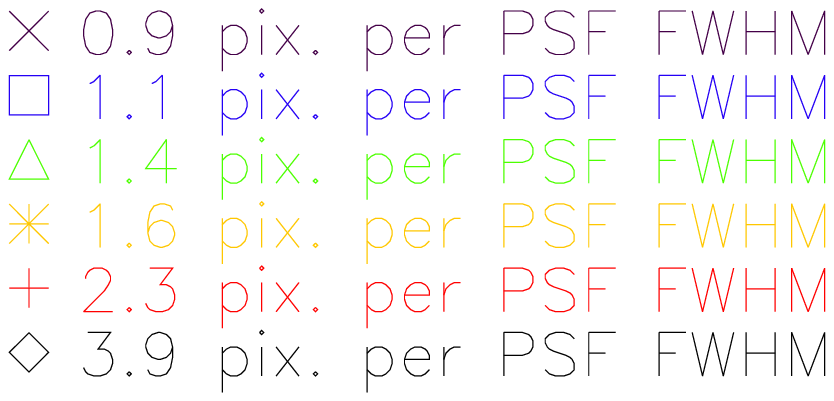}\\
\end{center}
\includegraphics[width=9cm]{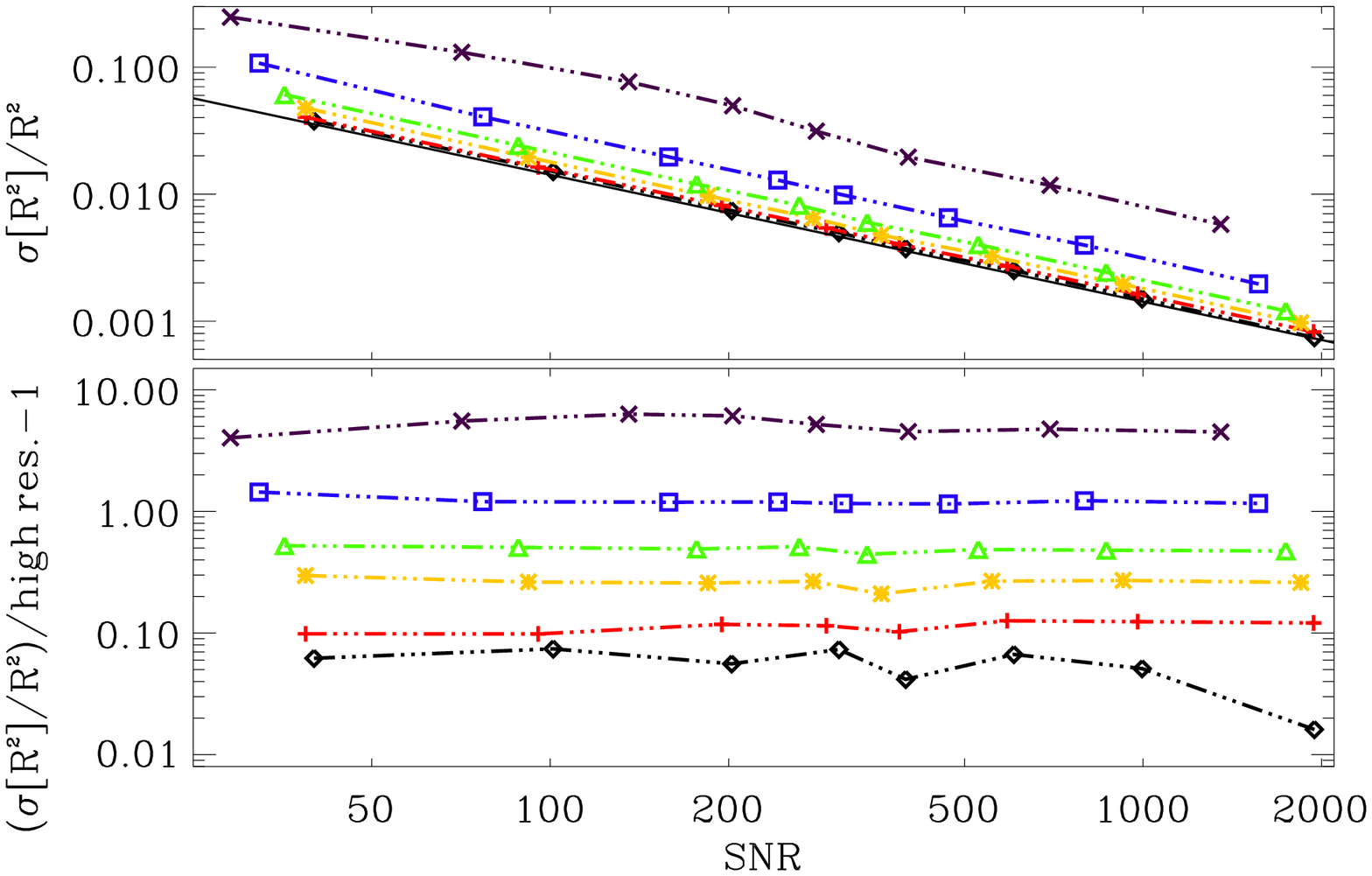}
\includegraphics[width=9.2cm]{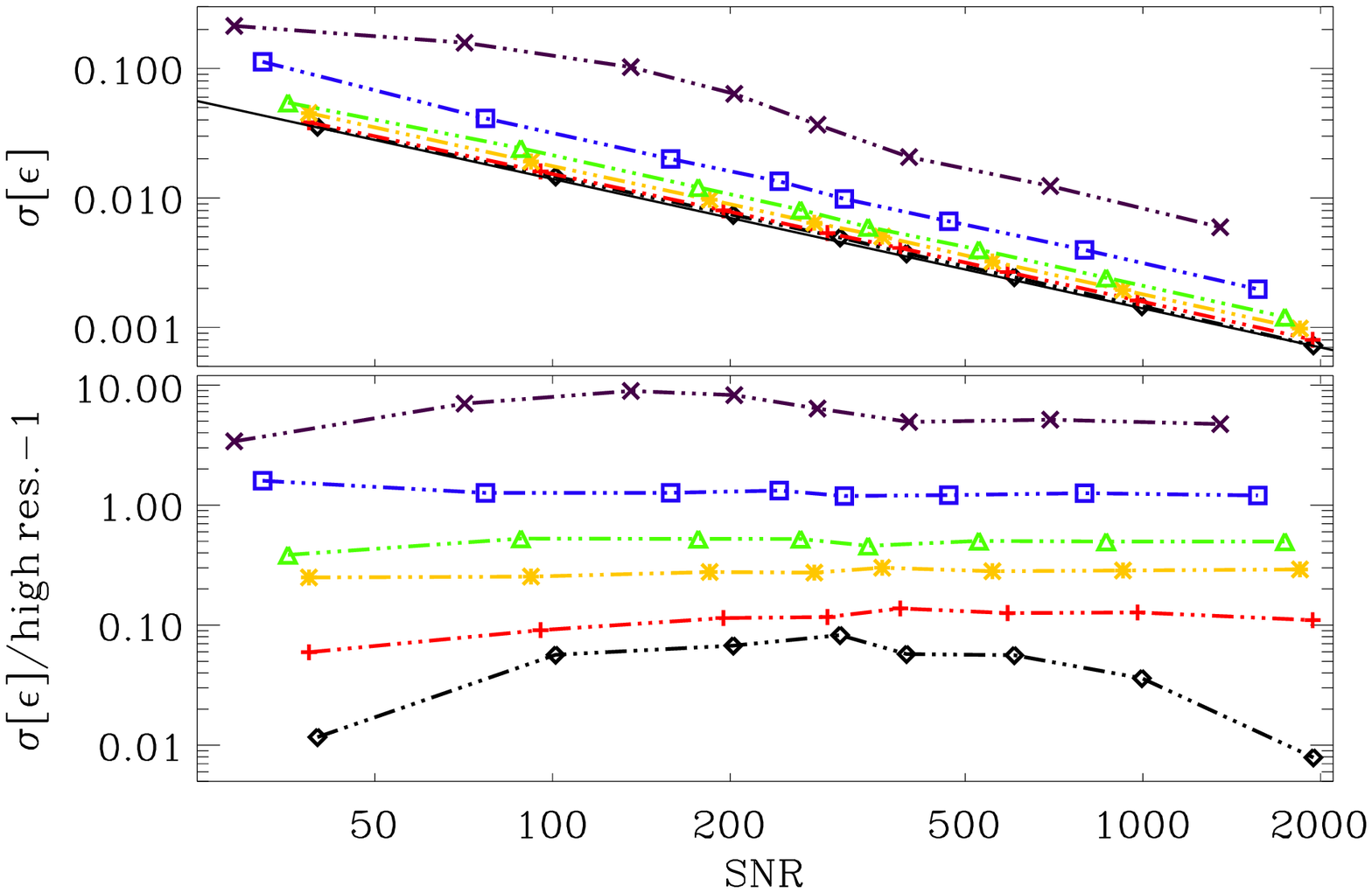}\\
\caption{
\label{fig:asymptoticlimits}
Solid lines: analytical predictions in the case of infinitly small pixels for $\sigma[\ellipticity]$ and $\sigma[\size^2]$ (see equations \ref{eq:analyticalgaussiansigmasize} and \ref{eq:analyticalgaussiansigmaell}). Data points joined by straight lines: simulation with finite pixel scales. All these results are for an elliptical gaussian PSF with an ellipticity $\ellipticity \simeq 0.1$. 
Each data point corresponds to about $10^4$ realisations. 
Black diamonds, red horizontal-vertical crosses, orange stars, green triangles, blue squares and purple diagonal crosses correspond to sizes $a_1=1.7$, 1, 0.7, 0.6, 0.5 and 0.4 pixels respectively. This corresponds to about $3.9$, $2.3$, 1.6, 1.4, 1.2 and 0.9 pixels respectively per PSF FWHM. 
Each data point corresponds to about $10^4$ realisations. 
On the upper panels the y-axis shows $\sigma[\size^2]/\size^2$ and $\sigma_\ellipticity$ , while on the bottom panels it shows the relative differences as compared to the infinite resolution case. The Signal-to-Noise Ratio is noted `SNR' rather than $\mathcal{S}$ in the text.
}
\end{figure*}

Figure \ref{fig:asymptoticlimits} 
shows the measured standard deviations on $\size^2$ and $\ellipticity$ as a function of $\snr$, for a range of pixel scales (from 0.9 to 3.9 pixels per PSF FWHM).
The black diamond symbols correpond to the highest resolution simulations (with 3.9 pixels per PSF FWHM). These points are close to the predictions (black curve) which are calculated analytically for the idealised case with infinitly small pixels (i.e. inifinitly high resolution, see section \ref{sec:ellgauss}). 
The upper and bottom panels show different representations of the same curves: on upper panels the y-axis shows $\sigma$, while on the bottom panels, it shows the relative differences with the infinite resolution case. 
As the pixel size increases (i.e. when the resolution decreases), the standard deviation increases monotonically and moves away from the analytical predictions. This shows that the analytic model can describe a high resolution case while simulations are required to take into account pixelation effects.

Our results in this paper are based on analytical predictions and assume the optimistic case where pixelation is good enough not to degrade the results.  From Figure \ref{fig:asymptoticlimits} we see that, for the smallest pixel scale we consider (3.9 pixels per PSF FWHM), the results from the analytic calculations and the simulations agree to better than 10\%. 
As the size of the pixels is increased to 0.9 pixels per 
PSF FWHM we see a dramatic increase 
of the standard deviation of a factor $\sim 10$. 
Some of this pixel scale effect will be mitigated through the use of dithering, but more simulations are needed to fully quantify the extent to which this will help.

%%%%%%%%%%%%%%%%%%%%%%%%%%%%%%%%%%%%%%%%%%%%%%%%%%%%%%%%%%
%%%%%%%%%%%%%%%%%%%%%%%%%%%%%%%%%%%%%%%%%%%%%%%%%%%%%%%%%%

\subsection{Shapelet PSF}
\label{sec:spacesteppsf}

We now turn our attention to the expected standard deviations of $\size^2$ and $\ellipticity$ (predicted by equations \ref{eq:sigsize2} and \ref{eq:sigell2}) for more complicated PSFs. 
We still work under the setup that the model fitted to the data can exactly describe the truth. With the notation used in section \ref{sec:general2dfit}, this means that when we fit the parameters $\paramvector$ of a model $m(\bs{x},\paramvector)$ there exists a solution $\paramvector({\rm PSF})$ that exactly describes the PSF.

\begin{figure*}
\includegraphics[bb=10 10 400 400,width=3.5cm]{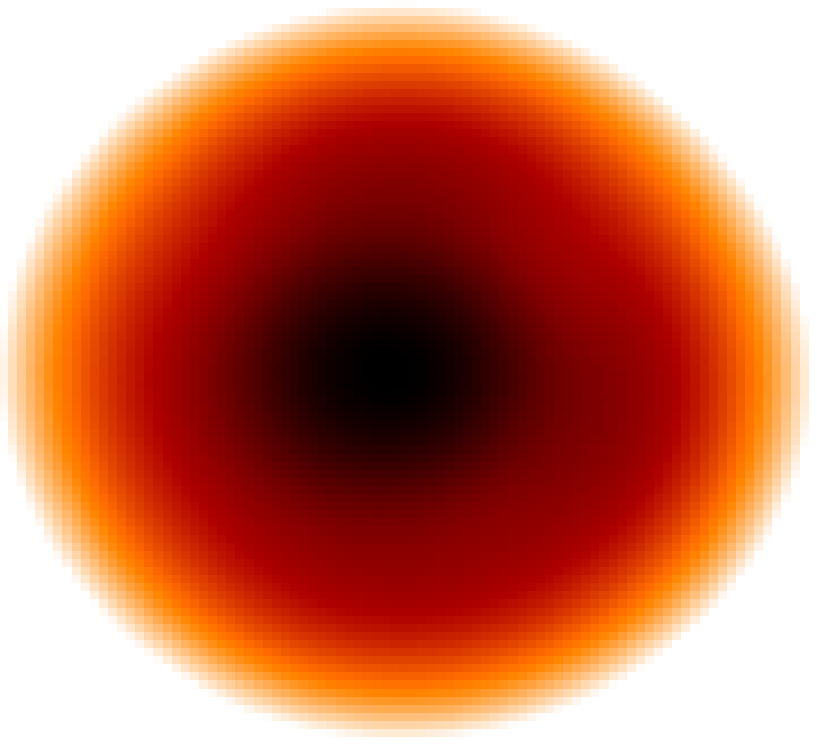}
\hspace*{-0.8cm}
\includegraphics[bb=10 10 400 400,width=3.5cm]{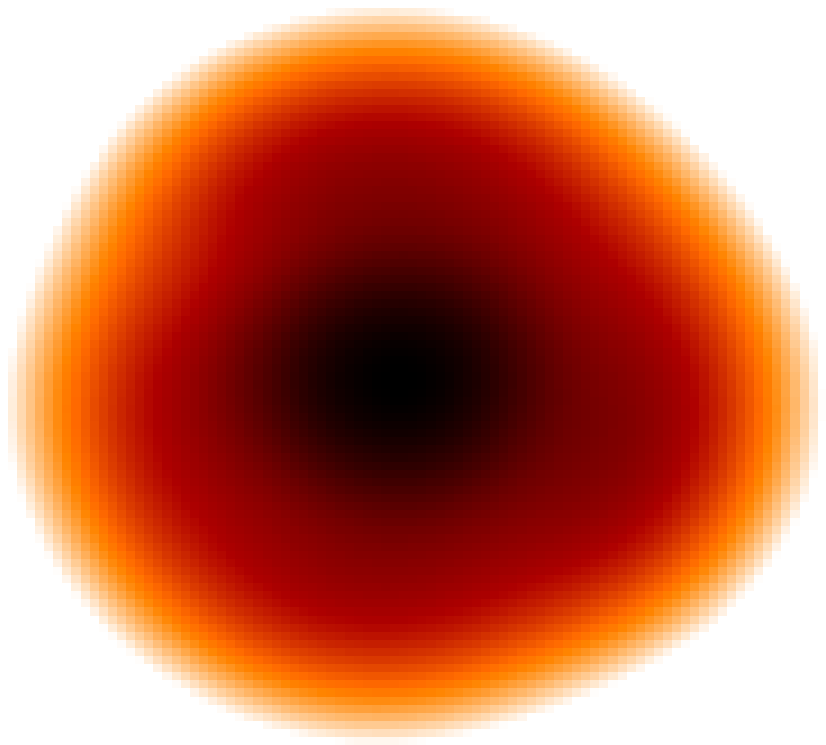}
\hspace*{-0.8cm}
\includegraphics[bb=10 10 400 400,width=3.5cm]{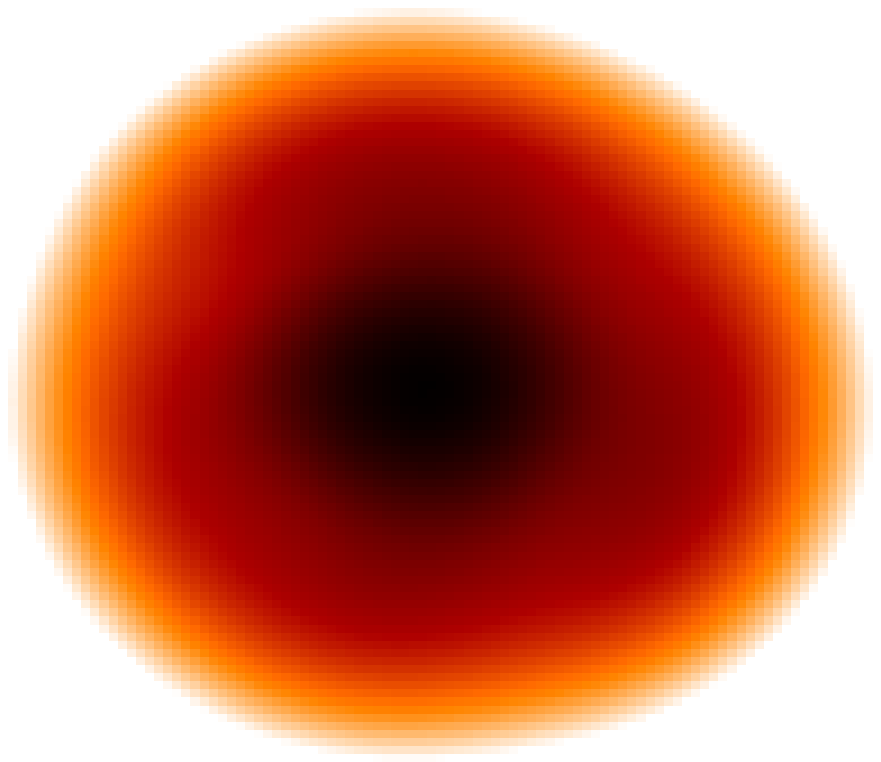}
\hspace*{-0.8cm}
\includegraphics[bb=10 10 400 400,width=3.5cm]{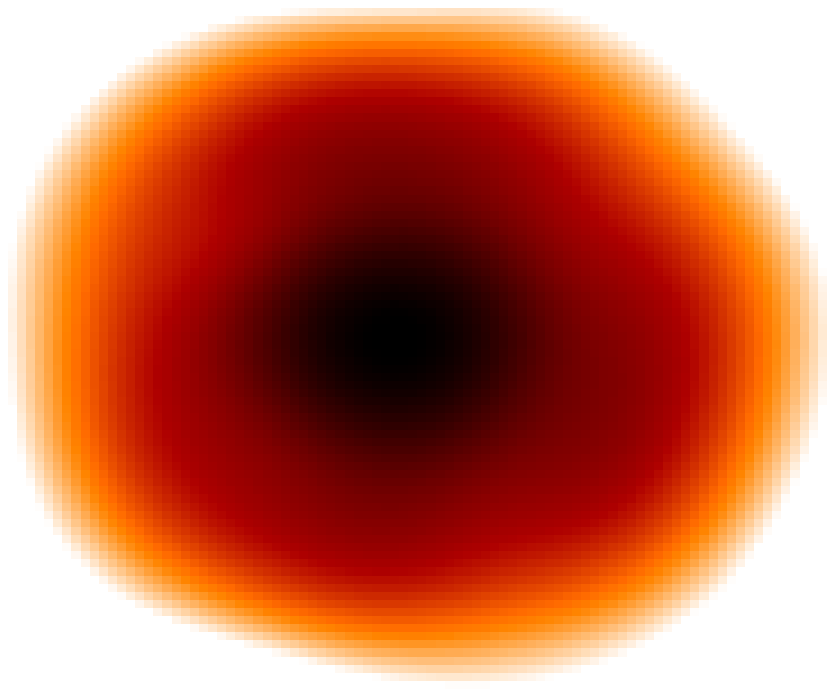}
\hspace*{-0.8cm}
\includegraphics[bb=10 10 400 400,width=3.5cm]{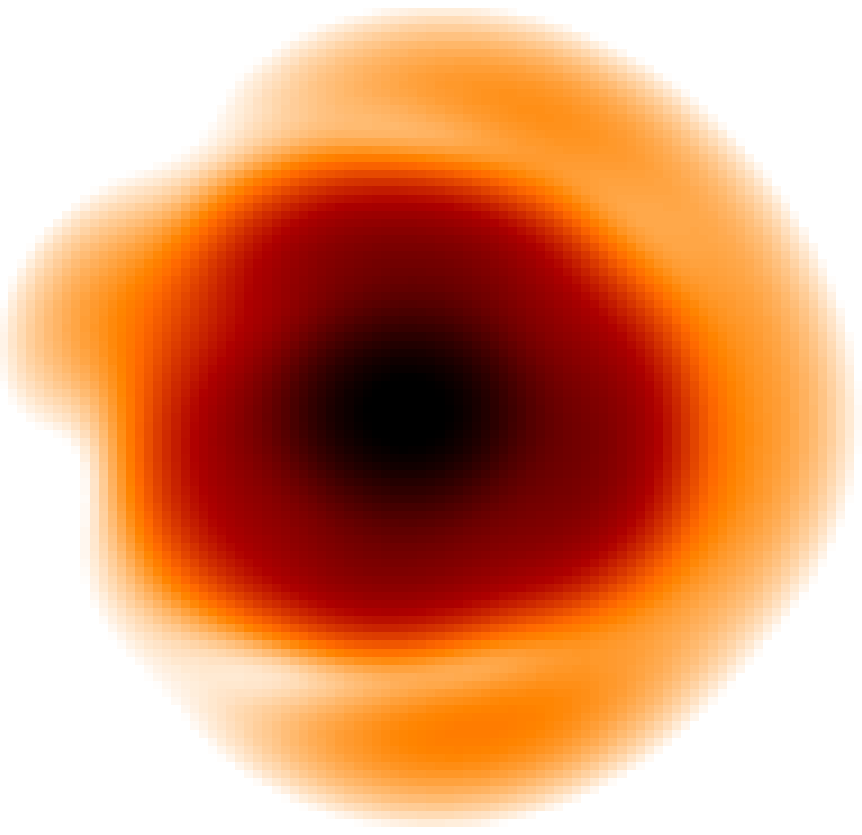}
\hspace*{-0.8cm}
\includegraphics[bb=10 10 400 400,width=3.5cm]{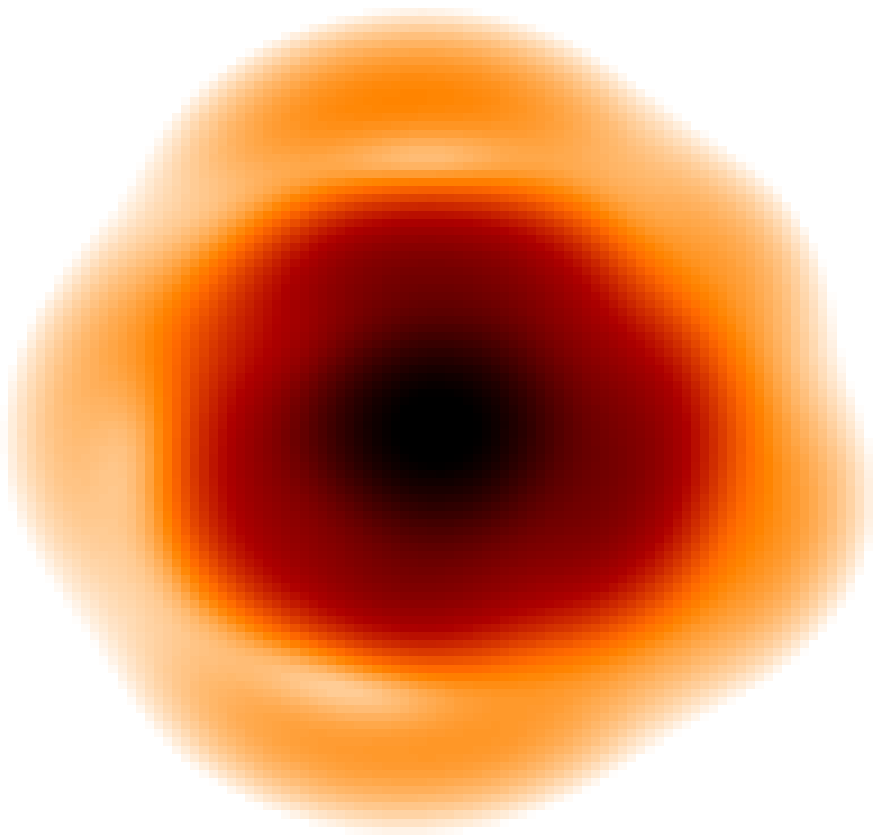}
\vspace*{-0.8cm}\\
\vspace*{-0.8cm}\\
\vspace*{-0.8cm}
\includegraphics[bb=10 10 400 400,width=3.5cm]{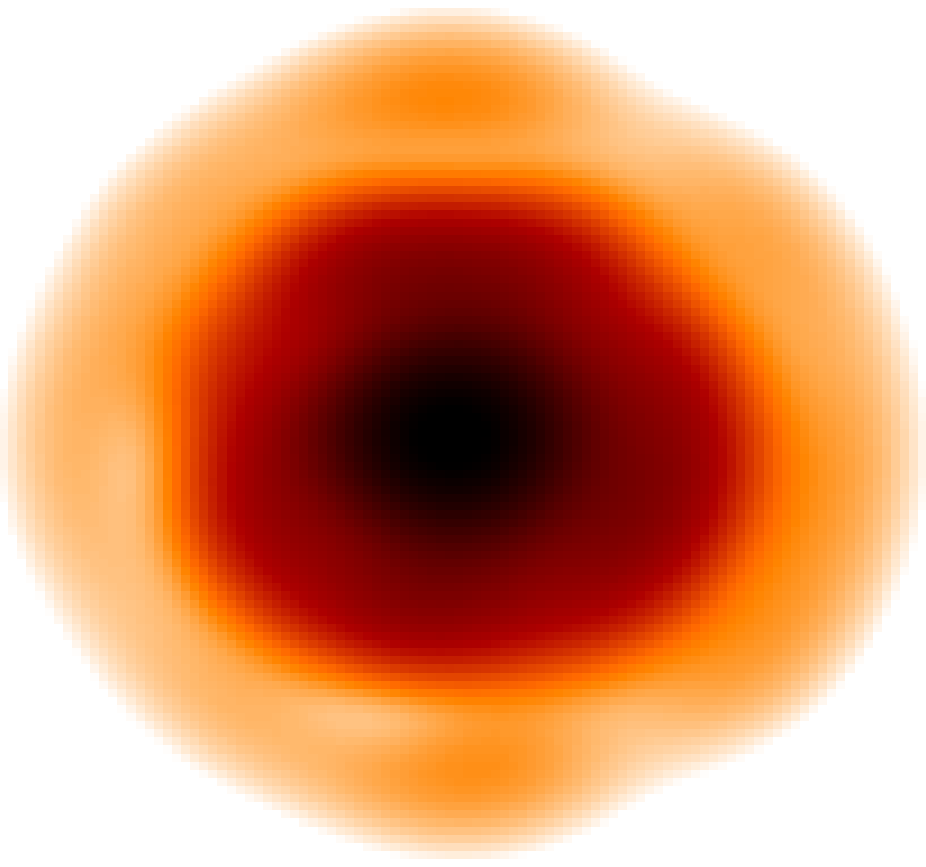}
\hspace*{-0.8cm}
\includegraphics[bb=10 10 400 400,width=3.5cm]{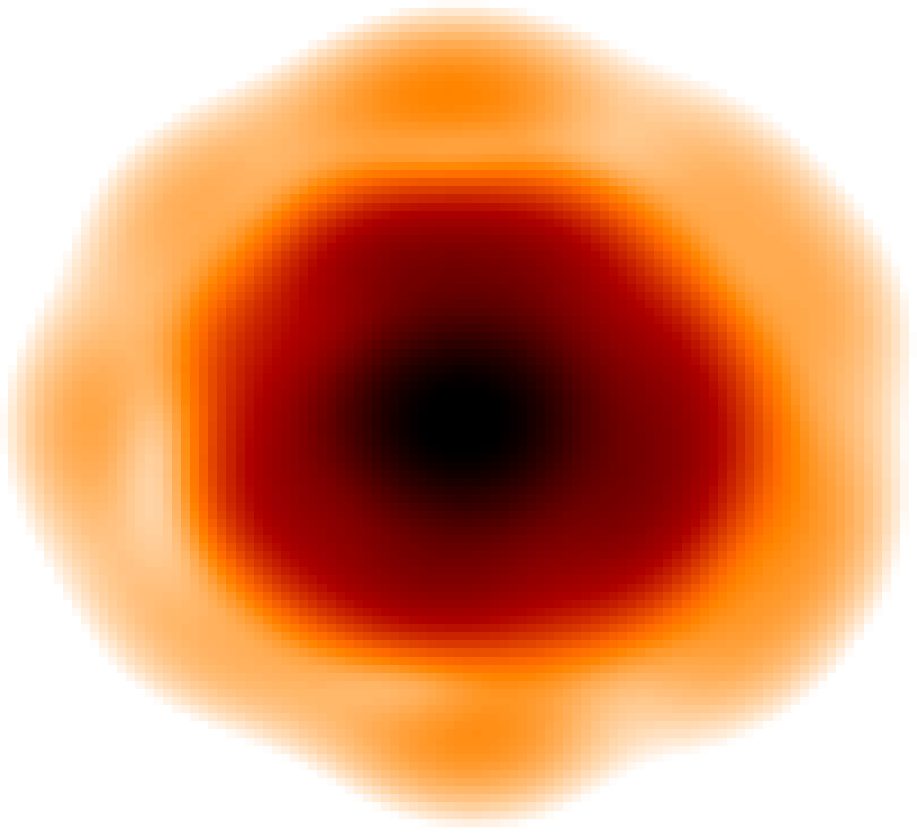}
\hspace*{-0.8cm}
\includegraphics[bb=10 10 400 400,width=3.5cm]{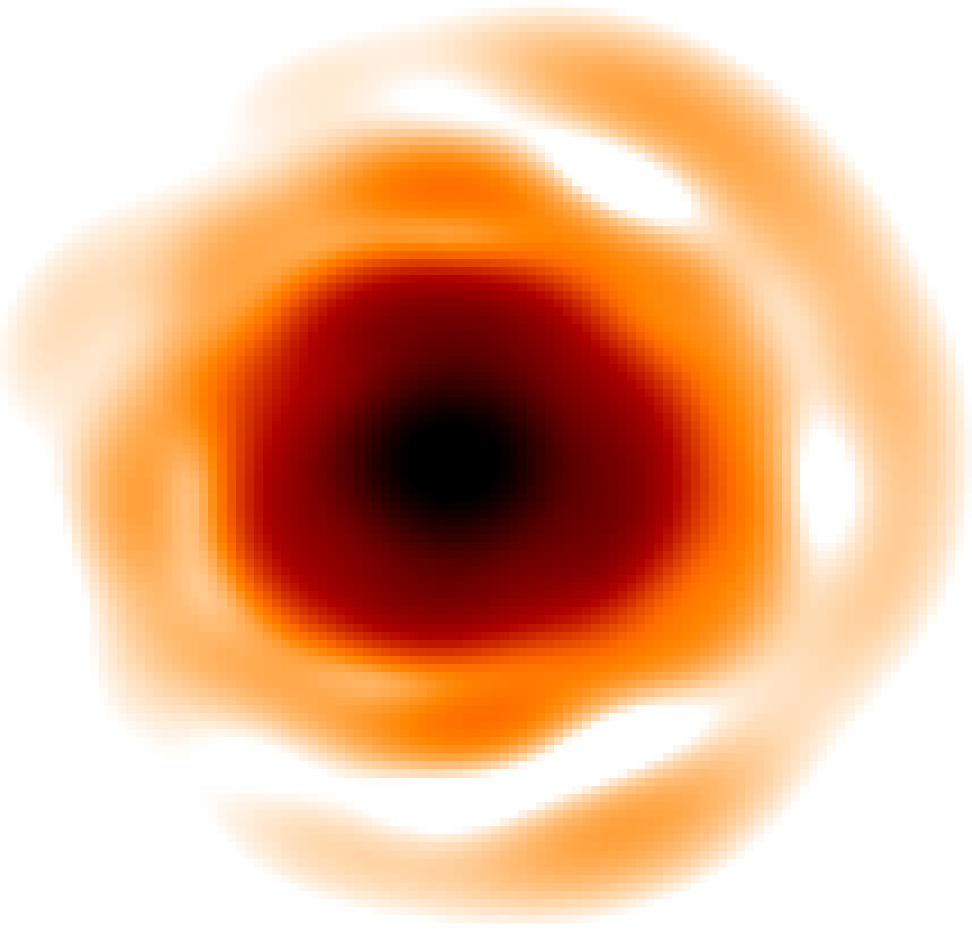}
\hspace*{-0.8cm}
\includegraphics[bb=10 10 400 400,width=3.5cm]{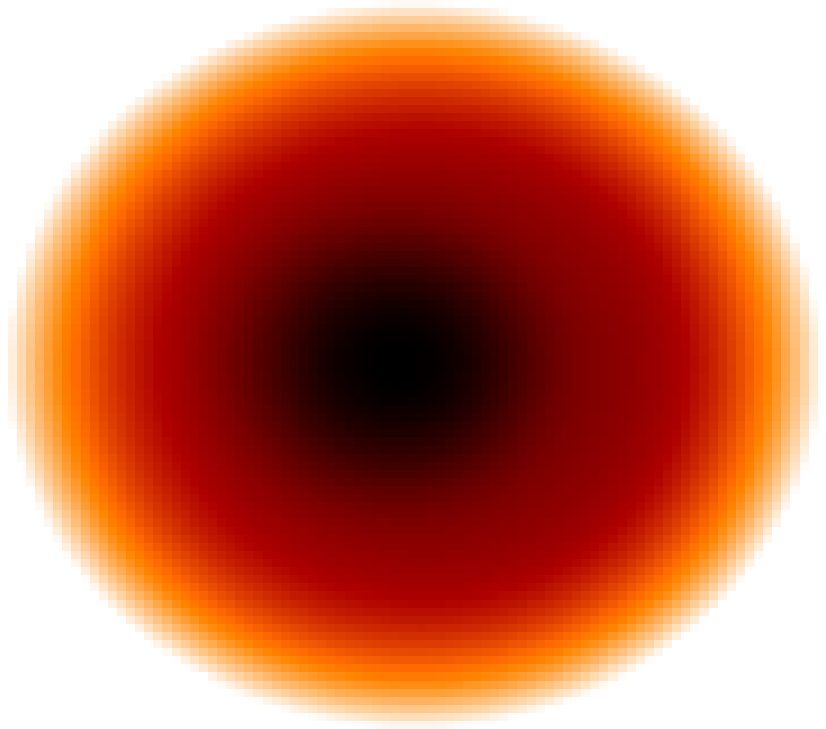}
\hspace*{-0.8cm}
\includegraphics[bb=10 10 400 400,width=3.5cm]{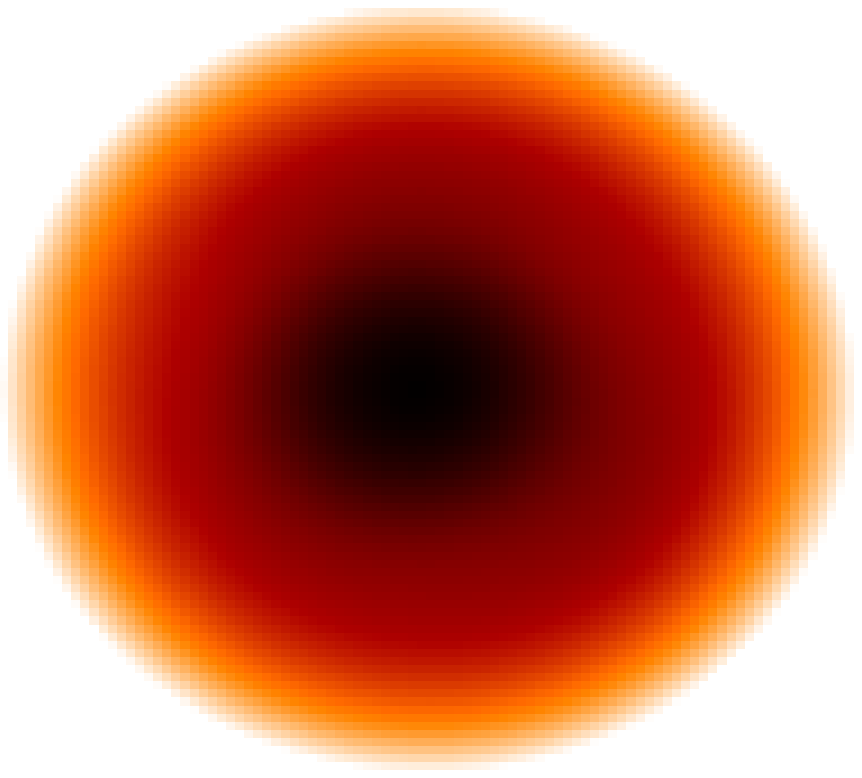}
\hspace*{-0.8cm}
\includegraphics[bb=10 10 400 400,width=3.5cm]{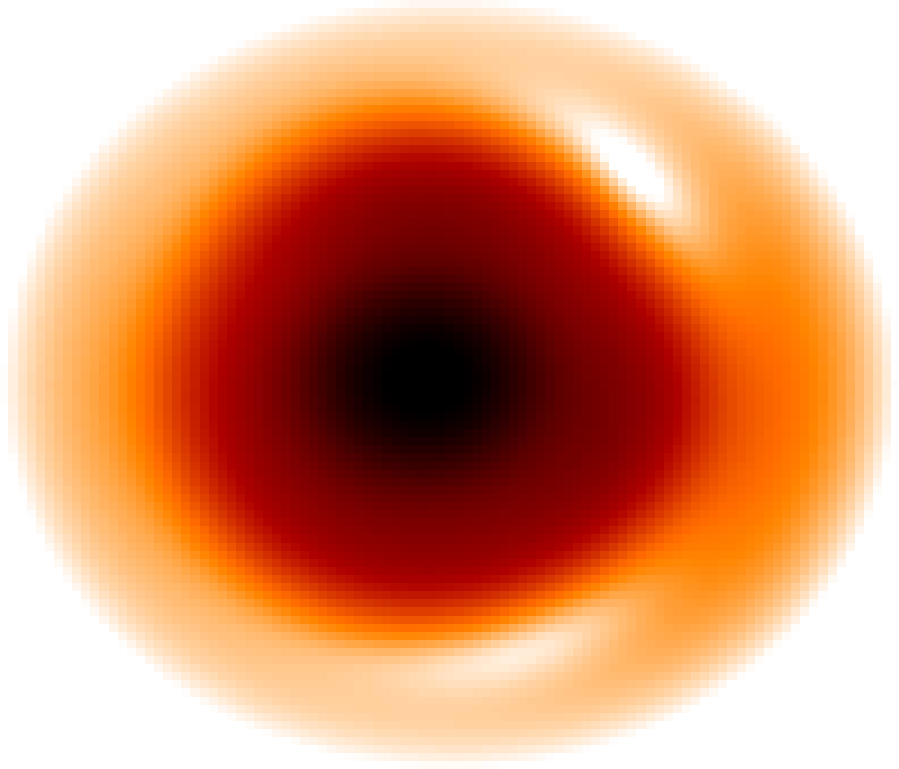}
\vspace*{-0.8cm}
\includegraphics[bb=10 10 400 400,width=3.5cm]{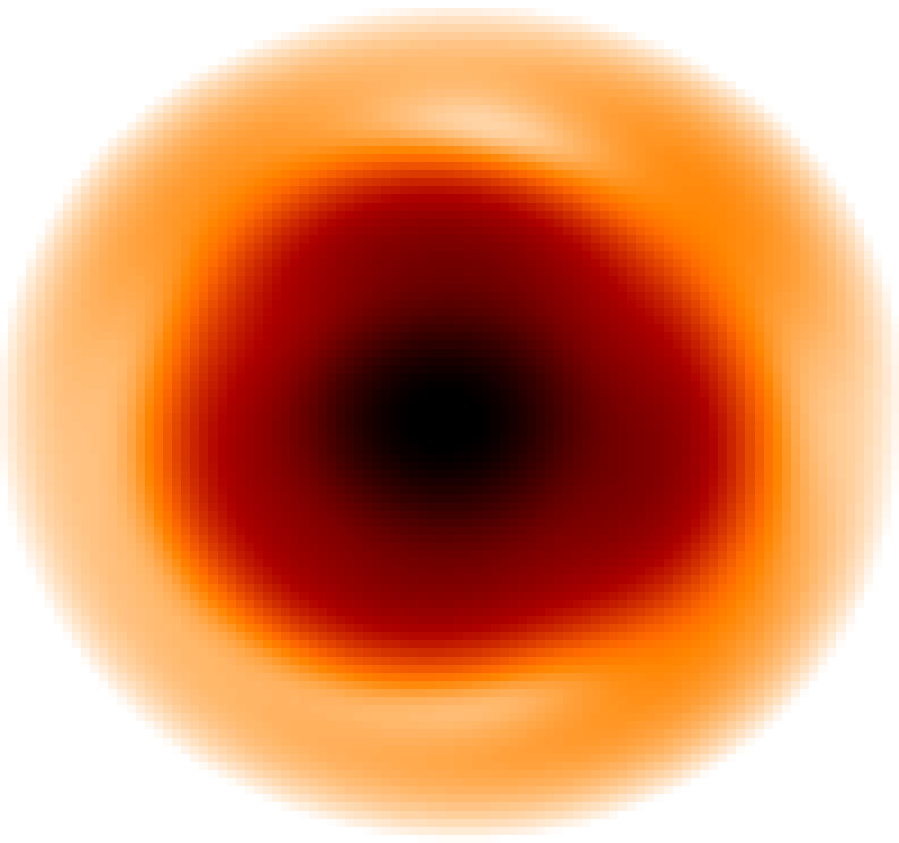}
\hspace*{-0.8cm}
\includegraphics[bb=10 10 400 400,width=3.5cm]{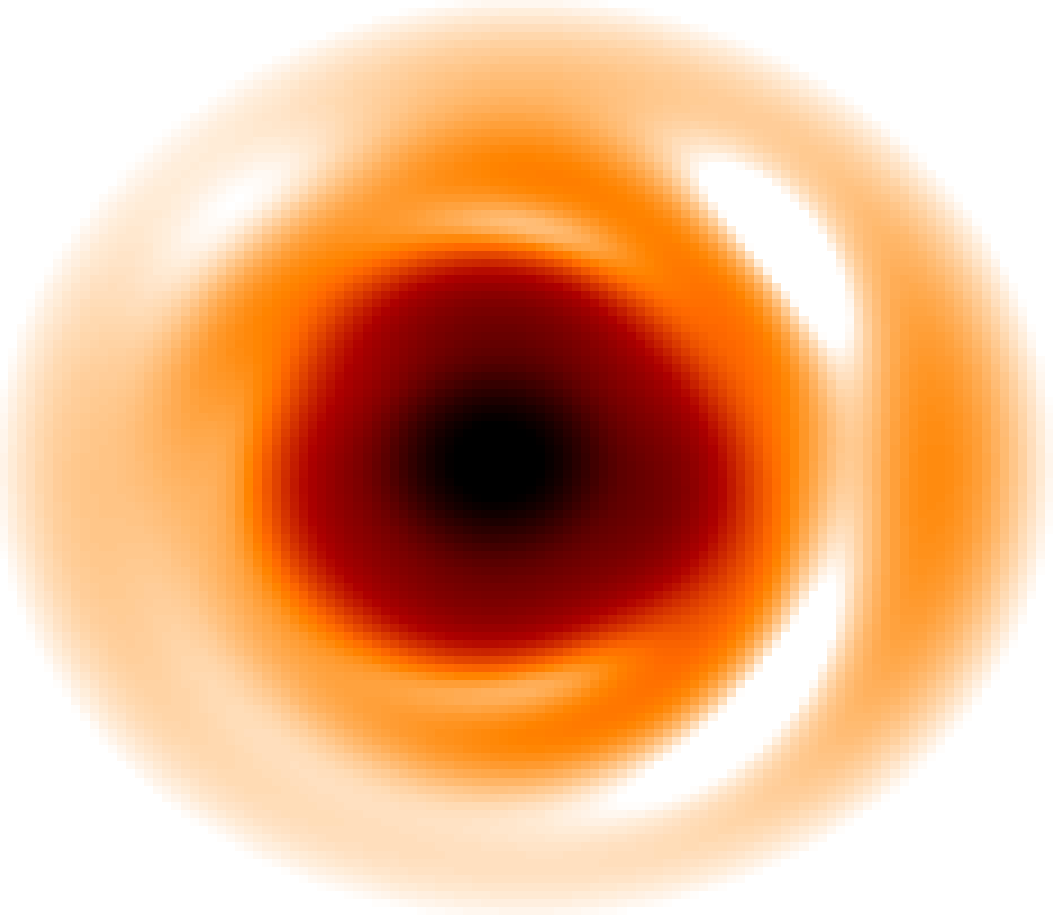}
\hspace*{-0.8cm}
\includegraphics[bb=10 10 400 400,width=3.5cm]{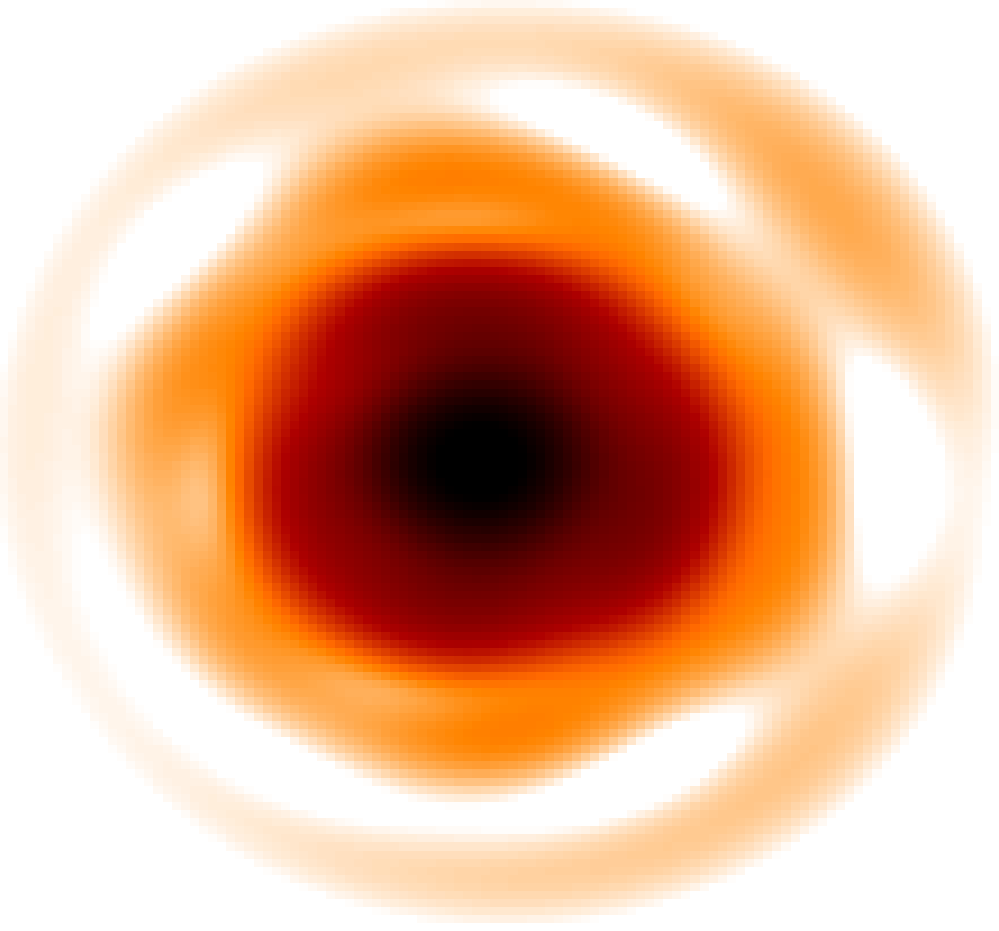}
\hspace*{-0.8cm}
\includegraphics[bb=10 10 400 400,width=3.5cm]{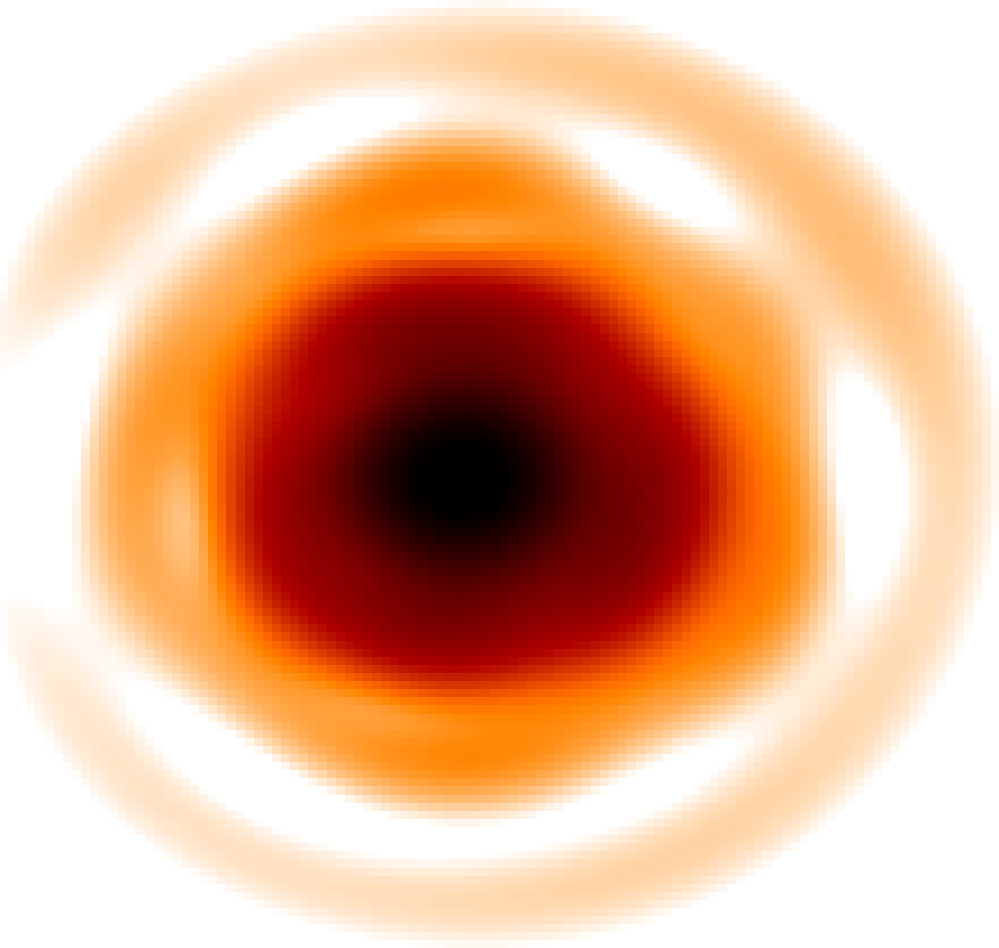}

\begin{center}
\includegraphics[width=15cm]{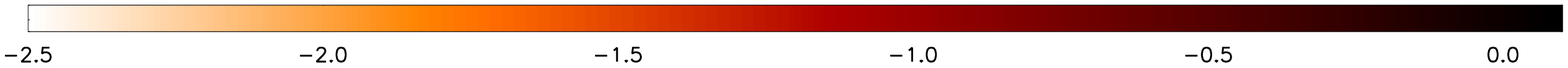}
\end{center}

\vspace*{-9.5cm}
\begin{tabular}{p{2.5cm}p{2.5cm}p{2.5cm}p{2.5cm}p{2.5cm}p{2.5cm}}
\hspace*{0.7cm}(a)\vspace*{2.2cm} &
\hspace*{0.7cm}(b) &
\hspace*{0.7cm}(c) &
\hspace*{0.7cm}(d) &
\hspace*{0.7cm}(e) &
\hspace*{0.7cm}(f)\\
\hspace*{0.7cm}(g)\vspace*{2.2cm} &
\hspace*{0.7cm}(h) &
\hspace*{0.7cm}(i) &
\hspace*{0.7cm}(j) &
\hspace*{0.7cm}(k) &
\hspace*{0.7cm}(l)\\
\hspace*{0.7cm}(m) &
\hspace*{0.7cm}(n) &
\hspace*{0.7cm}(o) &
\hspace*{0.7cm}(p) & & \\
\end{tabular}
\vspace*{3.3cm}
\caption{
\label{fig:psf}
Non-gaussian PSFs used to study the effect of complexity in section \ref{sec:spacesteppsf} (10 pixels per PSF FWHM, logarithmic color scale, the normalisation is such as the total flux is 10$^3$). Each of them is described with a different shapelet model, described in table \ref{tab:psf}.
}
\end{figure*}

We model the PSF by decomposing PSF-D of STEP III \citep{spacestep} into shapelets. 
We create 16 PSF models
with increasing levels of complexity by decomposing the STEP PSF
into basis sets with increasing $n_{\rm max}$ values. 
The fitted models are listed in table \ref{tab:psf}. 
This gives 16 slightly different PSFs close to \mbox{PSF-D} ($\ellipticity_i$ vary between -0.02 and 0.02) which are shown in figure \ref{fig:psf}.

The effect of the level of complexity of the PSF model on the
standard deviation of a measurement of the size and ellipticity
of the PSF is found by generating $\sim10^5$ realisations in the interval $10\le\snr\le1000$ for each of the 16 PSF shapelet basis sets and by fitting with the corresponding shapelet model.

\begin{figure*}
\resizebox{\hsize}{!}{
\includegraphics{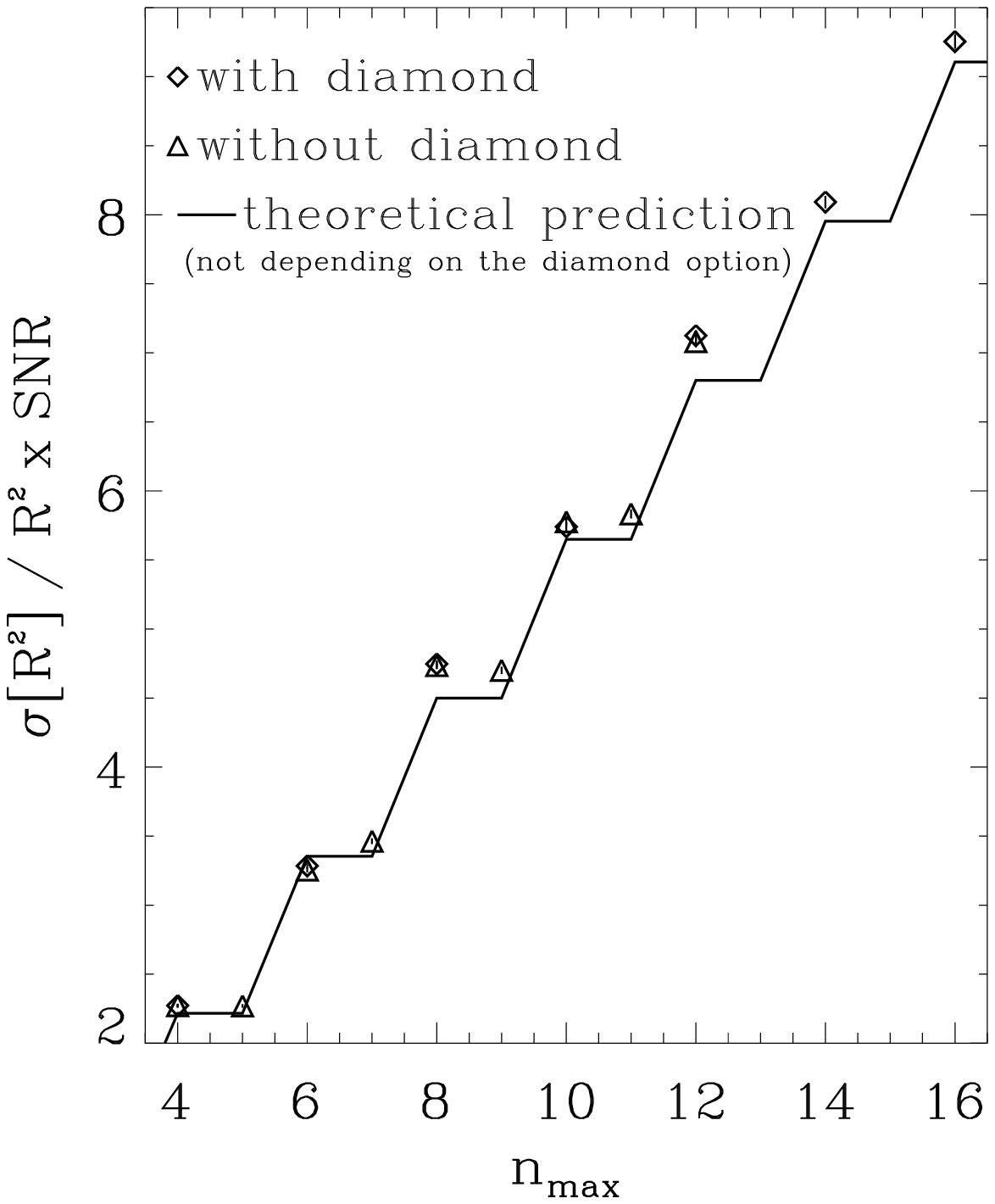}\\
\includegraphics{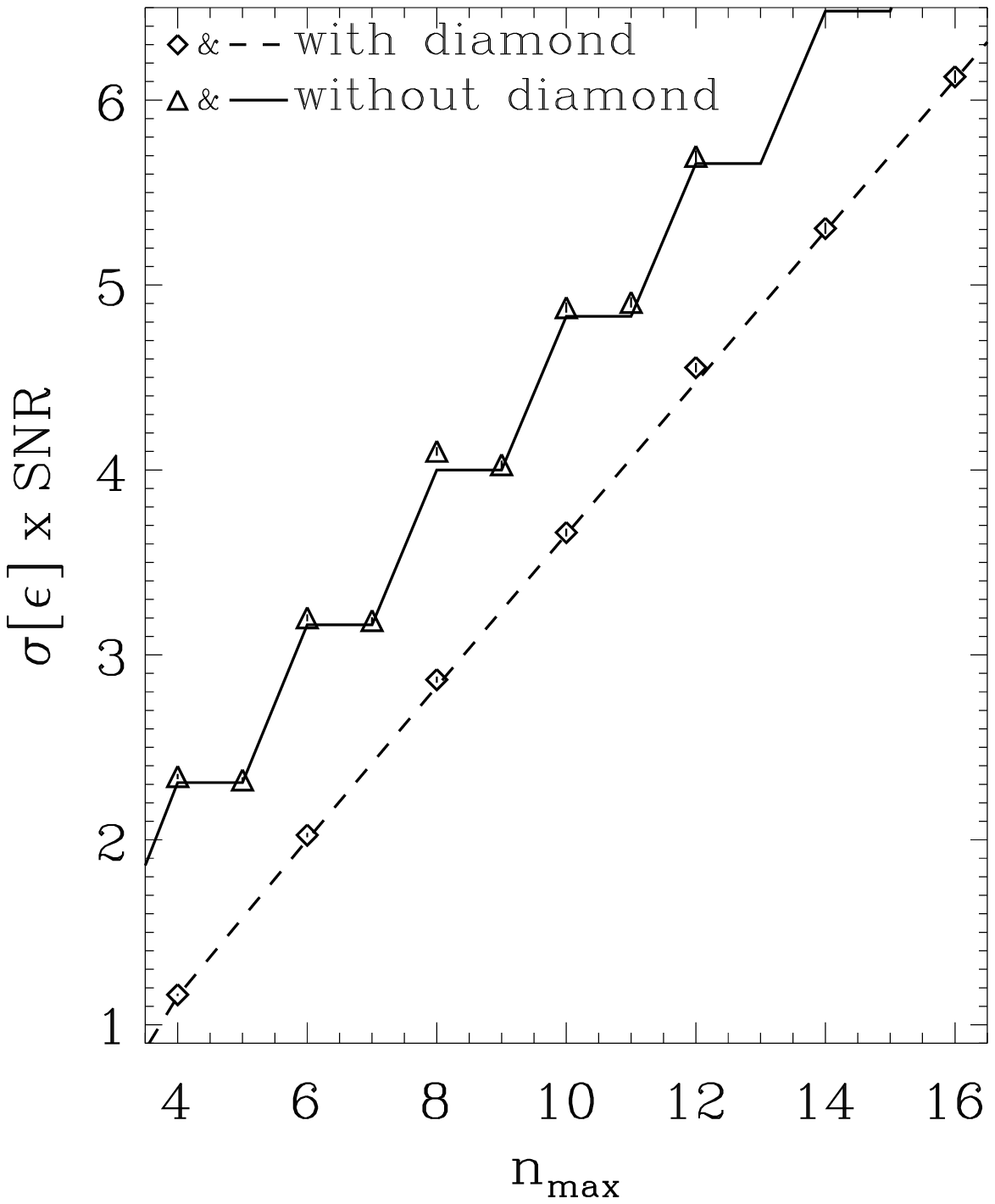}
}
\caption{
\label{fig:stddev.vs.nmax}
Lines: analytical predictions in the case of infinitely small pixels for $\sigma[\size^2]$ and $\sigma[\ellipticity]$ (see equations \ref{eq:sigsize2} and \ref{eq:sigell2}). 
Data points: simulation with small pixels (i.e. 10 pixels per PSF FWHM). 
All these results are for stars imaged through the complex PSFs shown in figure \ref{fig:psf} and 
with $10<\snr<1000$. 
Each data point corresponds to 
about $1.5\times 10^4$ realisations. 
Triangles and diamonds correspond to basis sets with and without the diamond configuration respectively. For $\sigma[\size^2]$ the solid line corresponds to the theoretical prediction (which does depend on the diamond option). For $\sigma[\ellipticity]$, solid and dashed lines correspond to theoretical predictions with and without the diamond option respectively. 
Without the diamond option, the complexity factors depend on the largest even integer lower than or equal to $n_{\rm max}$ (see equations \ref{eq:psidef1} and \ref{eq:psidef2}), this is why theoretical predictions look like a `staircase'. On the other hand this behaviour does not appear with the diamond 
configuration because the latter is defined only for even values of $n_{\rm max}$. We see that 
analytical predictions provide a good description, although they sometimes underestimate $\sigma[\size^2]$ and $\sigma[\ellipticity]$ by a few percent. 
The Signal-to-Noise Ratio is noted `SNR' rather than $\mathcal{S}$ in the text.
}
\end{figure*}

We compare the analytical predictions given by equations \ref{eq:sigsize2}, \ref{eq:sigell2} and \ref{eq:psidef1} to \ref{eq:psidef3} with simulations in figure \ref{fig:stddev.vs.nmax}. 
We see they provide a good description, although they sometimes underestimate $\sigma[\size^2]$ and $\sigma[\ellipticity]$ by a few percent. 
The results are similar whether or not the diamond 
configuration 
is used.

%%%%%%%%%%%%%%%%%%%%%%%%%%%%%%%%%%%%%%%%%%%%
%%%%%%%%%%%%%%%%%%%%%%%%%%%%%%%%%%%%%%%%%%%%
%%%%%%%%%%%%%%%%%%%%%%%%%%%%%%%%%%%%%%%%%%%%

\section{Requirements}
\label{sec:discussion}

In this section, we estimate the number of stars $N_*$ required to estimate the PSF to sufficient accuracy for a given survey. First, 
each star is an independent realisation of the PSF. Thus the PSF size and ellipticity can be estimated by:
\begin{eqnarray}
\size_{\rm PSF}^2 & = &\frac{\sum_k\size^2_{k}/\sigma^2[\size^2_{k}]}{\sum_k1/\sigma^2[\size^2_{k}]}
 = \frac{\sum_k\size^2_{k}\,\snr_{k}^2}{\sum_k\snr_{k}^2}\;,\\
\bs{\ellipticity}_{\rm PSF} & = &\frac{\sum_k\bs{\ellipticity}_{k}/\sigma^2[\ellipticity_{k}]}{\sum_k1/\sigma^2[\ellipticity_{k}]}
  = \frac{\sum_k\bs{\ellipticity}_{k}\,\snr_{k}^2}{\sum_k\snr_{k}^2}
\end{eqnarray}
where the sums are made on all the stars. These estimators are the minimum variance weighted averages if we assume the probability distribution in each of $\size^2$ and 
the two components of $\bs{\ellipticity}$ 
is a gaussian for each individual star. 
Second, 
variances are given by 
equations \ref{eq:sigsize2} and \ref{eq:sigell2} and can be written:
\begin{eqnarray}
\label{eq:sigell5}
\sigma^2[\size^2_{\rm PSF}] & = & 
\frac{\size_{\rm PSF}^4\,\psi_{\size^2}^2}{\sum_*{\snr}_{*i}^2}\equiv \frac{\size_{\rm PSF}^4\psi_{\size^2}^2}{N_*\,\snreff^2}\;,\\
\label{eq:sigell4}
\sigma^2[\ellipticity_{\rm PSF}] & = & \frac{\psi_\ellipticity^2}{\sum_*{\snr}_{*i}^2}\equiv \frac{\psi_\ellipticity^2}{N_*\,\snreff^2}\;.
\end{eqnarray}
We have defined an effective signal-to-noise of the stars $\snreff$ as the expected rms signal-to-noise given the maximum and minimum limiting values of $\snr$ used for PSF calibration ($\snrmax$ and $\snrmin$) and the number density of stars per unit of $\snr$: $dn / d\snr$, in the data set:
\be
\snreff^2 = \left(\int_{\snrmin}^{\snrmax} \frac{dn}{d\snr}\,d\snr\right)^{-1}\int_{\snrmin}^{\snrmax} \frac{dn}{d\snr}\,\snr^2\,d\snr\;.
\label{eq:eff_snr}
\ee
We then define the effective number of stars $N_*$ by
\be
N_*\,\snreff^2 = \sum_k\snr^2_{k}
\ee
where the sum is over the stars available around the galaxy. 
Therefore, the effective number of stars $N_*$ is equal to the actual number of stars if all the stars have the same $\snr$.

Substituting into equation \ref{eq:sigmasysdef} gives the requirement on the number of stars needed to calibrate the PSF. If the PSF ellipticity is at most a few percent then we 
can neglect the $\variance{\bs{\ellipticity}_{\rm PSF} }$ term leaving:
\be
\label{eq:reqnstar1}
N_* \gtrsim \frac{\left(P^\gamma\,\sigmasys^{\rm lim}\right)^{-2}}{\snreff^2}
      \left<\left(\frac{\size_{\rm PSF}}{\size_{\rm gal}}\right)^4\right>
 \left[
 \psi_{\size^2}^2\variance{\bs{\ellipticity}_{\rm gal}}+
 2\psi_\ellipticity^2\right]
\ee
where $\sigmasys^{\rm lim}$ is the upper limit acceptable for $\sigmasys$. 
The number of stars $N_*$ is a
central requirement of cosmic shear surveys: 
if we need to measure the PSF 
so that the systematic effects of the shear power
spectrum stay below $\sigmasys^{\rm lim.}$, we need to combine the
information from at least $N_*$ stars. Thus, for a given survey, one needs to estimate $N_*$ in
order to optimise the instrument and mission design. This is a requirement on the PSF stability. Indeed the star density must be taken into account to give the minimum area containing $N_*$ stars, over which the PSF must be stable.

Low $\snr$ stars do not contribute a great deal of information to the PSF
calibration. For instance if the density of stars scales roughly as
$dn/d\snr(\snr) \propto 1/\snr$, which is consistent with the simple star
count model of \cite{1980ApJ...238L..17B}, equation \ref{eq:eff_snr} shows that
the bright stars strongly dominate.  In fact, not only do low $\snr$ stars
bring little information, they may also induce a bias in the
PSF calibration.
This bias will be studied in forthcoming work. For the moment, we avoid this regime by adopting an arbitrary
lower limit of $\snrmin=100$.

On the other hand, the high $\snr$ cut off
$\snrmax$ and the star density distribution $dn/d\snr$, depend on
the properties of the surveys. They depend on a number of factors
including: the line of sight in the Milky Way, the instrumental
configuration and the observing strategy.
Equation \ref{eq:reqnstar1}, therefore, can not be computed for a general case, but it can be simplified and scaled to typical values, to give easily readable requirements on the star population in a data set. First, since the galaxy size distribution is steep, it is pessimistic but reasonable to approximate:
\be
\left<\left(\frac{\size_{\rm gal}}{\size_{\rm PSF}}\right)^4\right> \approx
\left[\left(\frac{\size_{\rm gal}}{\size_{\rm PSF}}\right)_{\rm min}\right]^{4}
\ee
where $(\size_{\rm gal}/\size_{\rm PSF})_{\rm min}$ is the minimum value that this ratio 
can reach, typically about 1.5. Second, we note that the variance $\variance{\bs{\ellipticity}_{\rm gal}}$ is around 0.16, as stated in section \ref{sec:systematics}, thus for usual PSF models where the complexity factor of the squared size $\psi_{\size^2}$ is of the same order as the complexity factor of the ellipticity $\psi_\ellipticity$ the expression inside the brackets in equation \ref{eq:reqnstar1} is driven by $\psi_\ellipticity^2$ and we can neglect $\psi_{\size^2}$. 
For example, for non-diamond shapelets with $n_{\rm max}=4$ or for diamond shapelets with $n_{\rm max}=6$, the ellipticity complexity factor is  
$\psi_\ellipticity\sim3$ 
(see table \ref{tab:psf}). 
Third, as mentioned in section \ref{sec:systematics}, \cite{2007arXiv0710.5171A} find for a DUNE-like cosmic shear experiment $(\sigmasys^{\rm lim.})^2=10^{-7}$. Fourth, we take for $\snreff$ a typical value of 500, 
which is roughly $\snr$ of an AB magnitude 20 star when imaged with a DUNE-like telescope (1500 seconds in a broad RIZ band). 
With these central values, we obtain:
\be
\label{eq:reqnstar2}
 N_* \! \gtrsim 50
   {\left(\frac{\snreff}{500}\right)\!\!}^{-2} \!   {\left(\frac{\left(\!\size_{\rm gal}/\size_{\rm PSF}\!\right)_{\rm min}\!\!}{1.5}\right)\!\!}^{-4} \!
   {\left(\!\frac{\psi_\ellipticity}{3}\!\right)\!}^2 \!
   {\left(\!\frac{(\sigmasys^{\rm lim})^2\!}{10^{-7}}\right)\!\!\!}^{-1}\;.
\ee
Therefore the PSF of a DUNE-like survey will need to be stable over a region containing about 50 stars.

The scaling relation given by equation \ref{eq:reqnstar2} allows one to study different survey configurations. This is illustrated by table \ref{tab:surveys} that gives $N_*$ for 6 typical configurations comparable to: the Canada-France-Hawai-Telescope Legacy Survey (CFHTLS), the Kilo-Degree Survey\footnote{{\tiny {\textsf http://www.eso.org/sci/observing/policies/cSurveys/sciencecSurveys.html}}} (KIDS), the  Large Synoptic Survey Telescope (LSST), the SuperNovae Acceleration Probe (SNAP) and the Dark UNiverse Explorer (DUNE). 
To build this table, we have estimated the accuracy $\sigma[w_0]$ and $\sigma[w_a]$ that can be achieved according to the surface covered, the median redshift $z_m$ and the galactic surface density $n_g$ in the case where systematics are lower than statistical errors and when fitting the 7 cosmological parameters: $\Omega_m$, $\Omega_b$, $\sigma_8$, $h$, $w_0$, $w_a$ and $n$. This estimation is made through the scaling relations proposed by \cite{2007MNRAS.381.1018A}. 
We have then estimated the value of $(\sigmasys^{\rm lim.})^2$ required to achieve this assumption of sub-dominant systematics, according to \cite{2007arXiv0710.5171A}. Finally we have computed $N_*$ according to our scaling relation (eq. \ref{eq:reqnstar2}).
\begin{enumerate}
\item The two first lines show surveys comparable to the current largest data sets optimised for cosmic shear. For instance the CFHTLS which currently covers $50\,$deg$^2$ \citep{2007arXiv0712.0884F} and will eventually cover $170\,$deg$^2$. The relatively poor constraints got on $w_0$ with such surveys (and the absence of constraint on $w_a$) illustrate the fact that cosmic shear surveys of the current generation do not constraint significantly the dark energy. They rather aim to constraint the combination $\sigma_8\times\Omega_M^\alpha$. However this does not change  the requirement on $\sigmasys^{\rm lim.}$ since \cite{2007arXiv0710.5171A} show that this requirement does not depend on the considered cosmological parameter. One can see that these surveys need to calibrate their PSF over few stars. Assuming a typical star density of $\sim 1.{\rm arcmin}^{-2}$ at these magnitudes, this shows that at scales smaller than few arcmin the PSF correction may introduce significant systematics. This is consistent with the results of \cite{2007MNRAS.381..702B} who describe a joint analysis of $100\,$deg$^2$ from several surveys and find significant B modes on these scales.
\item The third line corresponds to a cosmic shear survey of the next generation, such as KIDS/VIKING.
\item The fourth line shows an ambitious deep survey from the ground covering $15,000\,$deg$^2$, which is the largest cosmic shear survey that can be perfomed from Mauna Kea with a small airmass\footnote{On one hand, half of the sky is available for cosmic shear, corresponding to $\sim$20,000$\,$deg$^2$. The other half is masked by the Milky Way. This is why one uses to call `full-sky' the surveys covering 20,000$\,$deg$^2$. On the other hand, in order to maintain the air mass in a reasonable range, we make the optimistic assumption that the minimum elevation of ground observations is $50\,$deg in local horizon coordinates. With this constraint, a telescope located at Mauna Kea (i.e. Hawaii) can cover about $15,000\,$deg$^2$ during the year}. This illustrates one of the main limitation of doing cosmic shear from the ground: although the survey is rather deep, the galaxy surface density $n_g$ and the median redshift $z_m$ are limited. This is due to the fact that faint galaxies are too small: they have sizes comparable to the PSF and can not be included in a cosmic shear analysis. This is included in the scaling relation \ref{eq:reqnstar2} through the minimum dilution factor $(R_{\rm gal}/R_{\rm PSF})_{\rm min}$. Note that it is not possible to change significantly the reference value of $(R_{\rm gal}/R_{\rm PSF})_{\rm min}=1.5$ adopted in equation \ref{eq:reqnstar2} because it is at the power 4. For instance, a change from 1.5 to 1.2 to increase $n_g$ would also increase $N_*$ by a factor 2.5. 
Such a survey is similar to the LSST \citep{2006AAS...209.8619K} except that 
the LSST collaboration is developing a specific observing strategy that aims to drastically reduce $\psi_\ellipticity$. This strategy is based on stacking many very short exposures taken with different orientations of the camera in order to circularise the PSF. Another effect of this strategy is that the effective Signal-to-Noise Ration $\snreff$ of stars may be much larger than 300, as assumed in table \ref{tab:surveys}. But this strategy also has a fundamental limit: the more $\psi_\ellipticity$ is reduced, the more $\psi_{\size^2}$ increases (for instance because of the astrometric errors while stacking the low Signal-to-Noise Ratio exposures). For such a LSST-like strategy, estimating $N_*$ through equation \ref{eq:reqnstar2} is not relevant because in this equation we assume that $\psi_\ellipticity$ and $\psi_{\size^2}$ are of the same order. $N_*$ must be estimated through equation \ref{eq:reqnstar1}.
\item The fifth line corresponds to a deep space survey covering $1000\,$deg$^2$. This is similar to SNAP as described by \cite{2004AJ....127.3089M} and \cite{2004AJ....127.3102R}.
\item The last line corresponds to a DUNE-like survey (full sky from space with a medium depth and a reasonable galaxy density) adopted as the reference in the scaling relation (eq. \ref{eq:reqnstar2}).
\end{enumerate}
\begin{table*}
\begin{center}
\begin{tabular}{|c|c|c|c|c|c|c|}
\cline{2-7}
\multicolumn{1}{l|}{ } & surface & $z_m$ & $n_g$  & $\sigma[w_0]$ ($\sigma[w_a]$) & $(\sigmasys^{\rm lim.})^2$ & $N_*$\\
\multicolumn{1}{l|}{ } & (deg$^2$) &  & (.arcmin$^{-2}$) &  &  & $/ (\psi_\ellipticity/3)^2$\\
\cline{2-7}
\hline
Ground surveys & 50 & 0.8 & 10 & 0.9 (nr) & $4\times10^{-6}$ & 3\\
\cline{2-7}
& 170 & 0.8 & 10 & 0.5 (nr) & $2\times10^{-6}$ & 6\\
\cline{2-7}
($\mathcal{S}_{\rm eff}=300$) & 1,500 & 0.8 & 20 & 0.1 (0.6) & $6\times10^{-7}$ & 25\\
\cline{2-7}
 & 15,000 & 0.9 & 30 & 0.03 (0.1) & $10^{-7}$ & 104\\
\hline
Space surveys & 1,000 & 1.2 & 50 & 0.07 (0.3) & $3\times10^{-7}$ & 15\\
\cline{2-7}
($\mathcal{S}_{\rm eff}=500$) & \textbf{20,000 }& \textbf{0.9} & \textbf{40} & \textbf{0.02 (0.1)} & \textbf{10$^{-7}$} &  \textbf{50}\\
\hline
\end{tabular}
\caption{
\label{tab:surveys}
The required number of stars $N_*$ for PSF calibration of typical surveys, as predicted by our scaling relation (equation \ref{eq:reqnstar2}) in function of the effective Signal-to-Noise Ratio of stars $\snreff$ (defined in equation \ref{eq:eff_snr}) and of the scientific requirement given in terms of $(\sigmasys^{\rm lim})^2$ as defined by \cite{2007arXiv0710.5171A} (i.e. the upper limit on the variance of systematics to ensure that systematics are sub-dominant compared to statistical errors when fitting cosmological parameters). 
The estimations of $\sigma[w_0]$ and $\sigma[w_a]$ are made through the scaling relations proposed by \cite{2007MNRAS.381.1018A}. 
The mention `nr' means that the constraint is `not relevant'. Corresponding surveys (similar to the current generation of cosmic shear surveys) do not constraint the dark energy but rather constraint $\sigma_8$. However, $(\sigmasys^{\rm lim.})^2$ does not depend on the considered cosmological parameter and thus these poor constraints on $w_0$ do contain all the information to compute $(\sigmasys^{\rm lim.})^2$. 
We consider that stars have an average Signal-to-Noise Ratio of $500$ in space and $300$ from the ground. We also consider a fixed dilution factor $(R_{\rm gal}/R_{\rm PSF})_{\rm min}$ of 1.5 for every survey. 
The bold line corresponds to the set-up of a DUNE-like survey (i.e. full-sky from space with medium depth and galaxy density) as taken as the reference in the scaling relation. 
Note that the surface of 20,000$\,$deg$^2$ (i.e. half of the sky) is the maximum surface of the sky relevant for cosmic shear. That is why it is often called `full-sky'. In the same spirit, 15,000$\,$deg$^2$ is the maximum surface of the sky (relevant for cosmic shear) that can be covered from Mauna Kea with a reasonable airmass.
}
\end{center}
\end{table*}

%%%%%%%%%%%%%%%%%%%%%%%%%%%%%%%%%%%%%%%%%%%%%%%%%%%%%%
%%%%%%%%%%%%%%%%%%%%%%%%%%%%%%%%%%%%%%%%%%%%%%%%%%%%%%
%%%%%%%%%%%%%%%%%%%%%%%%%%%%%%%%%%%%%%%%%%%%%%%%%%%%%%

\section{Conclusions}
\label{eq:conclusion}

We have studied the PSF calibration requirements for cosmic shear to measure cosmological parameters. 
We connect the finite information that we are able to extract from stars to the statistical error of the PSF calibration, then to the error on the galaxy ellipticity estimation. We express our results in terms of the minimum number of stars ($N_*$) that are needed to calibrate the PSF in order to keep the systematic errors below statistical errors for cosmological parameter estimation. On scales smaller than the area containing $N_*$ stars there is not enough information coming from the stars to calibrate the PSF. Therefore, the systematic errors in the cosmological parameters may dominate over the statistical errors. This means that these small scales should not be used to constrain cosmology unless the variability is known to be extremely small. 
Our results show that for current cosmic shear surveys this scale is about an arcminute, which may explain residual systematics found on these scales in current analyses. In future all sky cosmic shear surveys, the data set will be increased by several orders of magnitude and a tight control of the PSF behaviour will be required to prevent this scale from being increased and to reach smaller scales. This places strong requirements on the hardware and observing strategy. For instance, for ground observations where the atmosphere prevents the PSF from being constant over several stars in a single exposure, 
\cite{2004astro.ph.12234J,2006JCAP...02..001J} suggest 
the shear correlation functions should be 
calculated by cross-correlating galaxies in different
exposures. The PSF calibration error contribution would then be uncorrelated and averaged down.

We also demonstrate how $N_*$ can depend on the complexity of the PSF. We define a `complexity factor' for two different shapelet parametrisations and an elliptical gaussian parametrisation. This complexity factor is a number lower than the number of parameters used to describe the PSF and increases with the PSF complexity.
There is a different complexity factor 
associated with each PSF shape parameter. 
For accurate galaxy shear measurements, we find that
the complexity factor for the PSF ellipticity is more important than that for the size. 
We summarise the dependence of $N_*$ as a function of complexity, star Signal-to-Noise Ratio, galaxy size and cosmological requirements in a convenient scaling relation (see equation \ref{eq:reqnstar2}). 
Note that we work under the setup that the PSF can be perfectly described by the model, thus we do not
include in our error budget 
the systematics due to a poor choice of the PSF model.

We find that pixelation degrades the PSF calibration accuracy. Our calculation of $N_*$ holds for the optimistic case of infinitely small pixels (i.e. infinitely high resolution). A finite pixel size increases $N_*$.  We have performed simulations to predict the increase in the case of an elliptical gaussian PSF. We show that for 2 pixels (or more) per FWHM, the rms scatter agrees with the standard deviation predicted for infinitely small pixels to within 10\%, which in turn can be translated into a 20\% increase in the required number of calibration stars.  For large pixels we see that the pixelation effect can be dramatic, for instance having 0.9 pixels per PSF FWHM would lead to a factor of 100 increase in the number of stars needed. This effect may be mitigated to some extent by using dithering, 
which is beyond the scope of this paper. The main results of this paper are for the high resolution case that provides a strict lower limit on $N_*$.

%%%%%%%%%%%%%%%%%%%%%%%%%%%%%%%%%%%%%%
%%%%%%%%%%%%%%%%%%%%%%%%%%%%%%%%%%%%%%
%%%%%%%%%%%%%%%%%%%%%%%%%%%%%%%%%%%%%%

\clearpage

\begin{table*}
\begin{center}
\begin{tabular}{|c|cc|ccc|}
\hline
 label & $n_{{\rm max}}$ & diamond & $N_{\rm coef}$ & $N_{\rm coef}^{m=2(0)}$ & complexity factor\\
 & & (yes or no) & & & $\Psi_\ellipticity$($\Psi_{\size^2}$)\\
\hline
\hline
 gauss & - & - & - & - & $\sqrt{2}(\sqrt{2})$\\
\hline
 a & 4 & no & 15 & 2(3) & 3.27(2.58)\\
 b & 5 & no & 21 & 2(3) & 3.27(2.58)\\
 c & 6 & no & 28 & 3(4) & 4.47(3.74)\\
 d & 7 & no & 36 & 3(4) & 4.47(3.74)\\
 e & 8 & no & 45 & 4(5) & 5.66(4.90)\\
 f & 9 & no & 55 & 4(5) & 5.66(4.90)\\
 g & 10 & no & 66 & 5(6) & 6.83(6.06)\\
 h & 11 & no & 78 & 5(6) & 6.83(6.06)\\
 i & 12 & no & 91 & 6(7) & 8.00(7.21)\\
\hline
 j & 4 & yes & 9 & 1(3) & 1.63(2.58)\\
 k & 6 & yes & 16 & 2(4) & 2.83(3.74)\\
 l & 8 & yes & 25 & 3(5) & 4.00(4.90)\\
 m & 10 & yes & 36 & 4(6) & 5.16(6.06)\\
 n & 12 & yes & 49 & 5(7) & 6.32(7.21)\\
 o & 14 & yes & 64 & 6(8) & 7.48(8.37)\\
 p & 16 & yes & 81 & 7(9) & 8.64(9.52)\\
\hline
\end{tabular}
\caption{
\label{tab:psf}
Properties of the PSF models investigated in section \ref{sec:spacesteppsf}. 
The first one is the elliptical gaussian (labelled `gauss'). 
For each of the 16 other basis sets, we study one PSF (labelled a to p) shown in figure \ref{fig:psf}. 
From left to right, the columns are: 
$n_{\textrm{max}}$ and diamond defined in section \ref{sec:shapelets} and by \cite{2005MNRAS.363..197M}; the number $N_{\rm coef}$ of coefficients in the basis; the numbers $N_{\rm coef}^{m=2(0)}$ of coefficients with $m=2$ (and $m=0$) in the basis and the complexity factors defined by equations \ref{eq:psidef1} to \ref{eq:psidef3}.
}
\end{center}
\end{table*}

%%%%%%%%%%%%%%%%%%%%%%%%%%%%%%%%%%%%%%

\section*{Acknowledgments}
We thank 
S\'ebastien Boulade and collaborators at EADS/Astrium for stimulating discussions which helped initiate this study. 
We acknowledge Jo\"el Berg\'e for his very convenient on-line manual ({\small \textsf{http://www.astro.caltech.edu/\~{ }jberge/shapelets/manual}}) about the shapelets IDL pipeline, which Richard Massey makes available on {\small \textsf{http://www.astro.caltech.edu/\~{ }rjm/shapelets}}.  
A. Amara would like to thank Prof. K. S. Cheng and Dr. T. Harko of Hong Kong University for their hospitality. S. Bridle acknowledges support from the Royal Society in the form of a University Research Fellowship. L. Voigt acknowledges support from the UK Science and Technology Facilities Council. 

%%%%%%%%%%%%%%%%%%%%%%%%%%%%%%%%%%%%%%
%%%%%%%%%%%%%%%%%%%%%%%%%%%%%%%%%%%%%%
%%%%%%%%%%%%%%%%%%%%%%%%%%%%%%%%%%%%%%

\bibliographystyle{aa}
\bibliography{bibfile}

\begin{thebibliography}{34}
\expandafter\ifx\csname natexlab\endcsname\relax\def\natexlab#1{#1}\fi

\bibitem[{{Albrecht} {et~al.}(2006){Albrecht}, {Bernstein}, {Cahn}, {Freedman},
  {Hewitt}, {Hu}, {Huth}, {Kamionkowski}, {Kolb}, {Knox}, {Mather}, {Staggs},
  \& {Suntzeff}}]{2006astro.ph..9591A}
{Albrecht}, A., {Bernstein}, G., {Cahn}, R., {et~al.} 2006, ArXiv Astrophysics
  e-prints

\bibitem[{{Amara} \& {Refregier}(2007{\natexlab{a}})}]{2007MNRAS.381.1018A}
{Amara}, A. \& {Refregier}, A. 2007{\natexlab{a}}, \mnras, 381, 1018

\bibitem[{{Amara} \& {Refregier}(2007{\natexlab{b}})}]{2007arXiv0710.5171A}
---. 2007{\natexlab{b}}, ArXiv e-prints, 710

\bibitem[{{Bahcall} \& {Soneira}(1980)}]{1980ApJ...238L..17B}
{Bahcall}, J.~N. \& {Soneira}, R.~M. 1980, \apjl, 238, L17

\bibitem[{{Bartelmann} \& {Schneider}(2001)}]{2001PhR...340..291B}
{Bartelmann}, M. \& {Schneider}, P. 2001, \physrep, 340, 291

\bibitem[{{Benjamin} {et~al.}(2007){Benjamin}, {Heymans}, {Semboloni}, {van
  Waerbeke}, {Hoekstra}, {Erben}, {Gladders}, {Hetterscheidt}, {Mellier}, \&
  {Yee}}]{2007MNRAS.381..702B}
{Benjamin}, J., {Heymans}, C., {Semboloni}, E., {et~al.} 2007, \mnras, 381, 702

\bibitem[{{Berg{\'e}} {et~al.}(2007){Berg{\'e}}, {Pacaud}, {R{\'e}fr{\'e}gier},
  {Massey}, {Pierre}, {Amara}, {Birkinshaw}, {Paulin-Henriksson}, {Smith}, \&
  {Willis}}]{2007arXiv0712.3293B}
{Berg{\'e}}, J., {Pacaud}, F., {R{\'e}fr{\'e}gier}, A., {et~al.} 2007, ArXiv
  e-prints, 712

\bibitem[{{Bernstein} \& {Jarvis}(2002)}]{2002AJ....123..583B}
{Bernstein}, G.~M. \& {Jarvis}, M. 2002, \aj, 123, 583

\bibitem[{{Fu} {et~al.}(2007){Fu}, {Semboloni}, {Hoekstra}, {Kilbinger}, {van
  Waerbeke}, {Tereno}, {Mellier}, {Heymans}, {Coupon}, {Benabed}, {Benjamin},
  {Bertin}, {Dor{\'e}}, {Hudson}, {Ilbert}, {Maoli}, {Marmo}, {McCracken}, \&
  {M{\'e}nard}}]{2007arXiv0712.0884F}
{Fu}, L., {Semboloni}, E., {Hoekstra}, H., {et~al.} 2007, ArXiv e-prints, 712

\bibitem[{{Heavens} {et~al.}(2006){Heavens}, {Kitching}, \&
  {Taylor}}]{2006MNRAS.373..105H}
{Heavens}, A.~F., {Kitching}, T.~D., \& {Taylor}, A.~N. 2006, \mnras, 373, 105

\bibitem[{{Heymans} {et~al.}(2006){Heymans}, {VanWaerbeke}, {Bacon}, {Berge},
  {Bernstein}, {Bertin}, \& {Bridle}}]{2006MNRAS.368.1323H}
{Heymans}, C., {VanWaerbeke}, L., {Bacon}, D., {et~al.} 2006, \mnras, 139, 313

\bibitem[{{High} {et~al.}(2007){High}, {Rhodes}, {Massey}, \&
  {Ellis}}]{2007astro.ph..3471H}
{High}, F.~W., {Rhodes}, J., {Massey}, R., \& {Ellis}, R. 2007, ArXiv
  Astrophysics e-prints

\bibitem[{{Hoekstra}(2003)}]{hoekstra03}
{Hoekstra}, H. 2003, ArXiv Astrophysics e-prints

\bibitem[{{Hoekstra} {et~al.}(1998){Hoekstra}, {Franx}, {Kuijken}, \&
  {Squires}}]{1998ApJ...504..636H}
{Hoekstra}, H., {Franx}, M., {Kuijken}, K., \& {Squires}, G. 1998, \apj, 504,
  636

\bibitem[{{Huterer} {et~al.}(2006){Huterer}, {Takada}, {Bernstein}, \&
  {Jain}}]{2006MNRAS.366..101H}
{Huterer}, D., {Takada}, M., {Bernstein}, G., \& {Jain}, B. 2006, \mnras, 366,
  101

\bibitem[{{Jain} {et~al.}(2006){Jain}, {Jarvis}, \&
  {Bernstein}}]{2006JCAP...02..001J}
{Jain}, B., {Jarvis}, M., \& {Bernstein}, G. 2006, Journal of Cosmology and
  Astro-Particle Physics, 2, 1

\bibitem[{{Jarvis} \& {Jain}(2004)}]{2004astro.ph.12234J}
{Jarvis}, M. \& {Jain}, B. 2004, ArXiv Astrophysics e-prints

\bibitem[{{Kahn} \& {LSST Collaboration}(2006)}]{2006AAS...209.8619K}
{Kahn}, S. \& {LSST Collaboration}. 2006, in Bulletin of the American
  Astronomical Society, Vol.~38, Bulletin of the American Astronomical Society,
  1020--+

\bibitem[{{Kaiser} {et~al.}(1995){Kaiser}, {Squires}, \&
  {Broadhurst}}]{1995ApJ...449..460K}
{Kaiser}, N., {Squires}, G., \& {Broadhurst}, T. 1995, \apj, 449, 460

\bibitem[{{Kuijken}(1999)}]{1999A&A...352..355K}
{Kuijken}, K. 1999, \aap, 352, 355

\bibitem[{{Luppino} \& {Kaiser}(1997)}]{1997ApJ...475...20L}
{Luppino}, G.~A. \& {Kaiser}, N. 1997, \apj, 475, 20

\bibitem[{{Massey} {et~al.}(2007){Massey}, {Heymans}, {Berg{\'e}}, {Bernstein},
  {Bridle}, {Clowe}, {Dahle}, {Ellis}, {Erben}, {Hetterscheidt}, {High},
  {Hirata}, {Hoekstra}, {Hudelot}, {Jarvis}, {Johnston}, {Kuijken},
  {Margoniner}, {Mandelbaum}, {Mellier}, {Nakajima}, {Paulin-Henriksson},
  {Peeples}, {Roat}, {Refregier}, {Rhodes}, {Schrabback}, {Schirmer}, {Seljak},
  {Semboloni}, \& {van Waerbeke}}]{2007MNRAS.376...13M}
{Massey}, R., {Heymans}, C., {Berg{\'e}}, J., {et~al.} 2007, \mnras, 376, 13

\bibitem[{{Massey} \& {Refregier}(2005)}]{2005MNRAS.363..197M}
{Massey}, R. \& {Refregier}, A. 2005, \mnras, 363, 197

\bibitem[{{Massey} {et~al.}(2004){Massey}, {Rhodes}, {Refregier}, {Albert},
  {Bacon}, {Bernstein}, {Ellis}, {Jain}, {McKay}, {Perlmutter}, \&
  {Taylor}}]{2004AJ....127.3089M}
{Massey}, R., {Rhodes}, J., {Refregier}, A., {et~al.} 2004, \aj, 127, 3089

\bibitem[{{Munshi} {et~al.}(2006){Munshi}, {Valageas}, {Van Waerbeke}, \&
  {Heavens}}]{munshi06}
{Munshi}, D., {Valageas}, P., {Van Waerbeke}, L., \& {Heavens}, A. 2006, ArXiv
  Astrophysics e-prints

\bibitem[{{Peacock} \& {Schneider}(2006)}]{2006Msngr.125...48P}
{Peacock}, J. \& {Schneider}, P. 2006, The Messenger, 125, 48

\bibitem[{{Refregier}(2003{\natexlab{a}})}]{2003MNRAS.338...35R}
{Refregier}, A. 2003{\natexlab{a}}, \mnras, 338, 35

\bibitem[{{Refregier}(2003{\natexlab{b}})}]{refregier03}
---. 2003{\natexlab{b}}, \araa, 41, 645

\bibitem[{{Refregier} \& {Bacon}(2003)}]{2003MNRAS.338...48R}
{Refregier}, A. \& {Bacon}, D. 2003, \mnras, 338, 48

\bibitem[{{Refregier} {et~al.}(2002){Refregier}, {Chang}, \&
  {Bacon}}]{2002sgdh.conf...29R}
{Refregier}, A., {Chang}, T.-C., \& {Bacon}, D.~J. 2002, in The shapes of
  galaxies and their dark halos, Proceedings of the Yale Cosmology Workshop
  ''The Shapes of Galaxies and Their Dark Matter Halos'', New Haven,
  Connecticut, USA, 28-30 May 2001. Edited by Priyamvada Natarajan. Singapore:
  World Scientific, 2002, ISBN 9810248482, p.29, ed. P.~{Natarajan}, 29--+

\bibitem[{{Refregier} {et~al.}(2004){Refregier}, {Massey}, {Rhodes}, {Ellis},
  {Albert}, {Bacon}, {Bernstein}, {McKay}, \&
  {Perlmutter}}]{2004AJ....127.3102R}
{Refregier}, A., {Massey}, R., {Rhodes}, J., {et~al.} 2004, \aj, 127, 3102

\bibitem[{{Rhodes} {et~al.}(2008){Rhodes}, {}, {}, {}, {}, {}, {}, \&
  {}}]{spacestep}
{Rhodes}, J., {}, {}, {et~al.} 2008, in prep.

\bibitem[{{Stabenau} {et~al.}(2007){Stabenau}, {Jain}, {Bernstein}, \&
  {Lampton}}]{2007arXiv0710.3355S}
{Stabenau}, H.~F., {Jain}, B., {Bernstein}, G., \& {Lampton}, M. 2007, ArXiv
  e-prints, 710

\bibitem[{{Voigt} \& {Bridle}(2008)}]{voigtb08}
{Voigt}, L. \& {Bridle}, S.~L. 2008, in prep

\end{thebibliography}

%%%%%%%%%%%%%%%%%%%%%%%%%%%%%%%%%%%%%%
%%%%%%%%%%%%%%%%%%%%%%%%%%%%%%%%%%%%%%
%%%%%%%%%%%%%%%%%%%%%%%%%%%%%%%%%%%%%%

\appendix

%%%%%%%%%%%%%%%%%%%%%%%%%%%%%%%%%%
%%%%%%%%%%%%%%%%%%%%%%%%%%%%%%%%%
%%%%%%%%%%%%%%%%%%%%%%%%%%%%%%%%%%

\section{Propagation of the PSF error}
\label{sec:propagation}
In this appendix we detail how we derive equations \ref{eq:errorpropagationpsfintogal} and \ref{eq:sigmasysdef} 
by propagating the error on each measurement of the PSF ellipticity and size to the error on the galaxy ellipticity and then to $\sigmasys$.

The 2 component galaxy ellipticity $\bs{\ellipticity}_{\rm gal}$ can be written in terms of radii $\size^2_{\rm obs}$ and $\size^2_{\rm PSF}$ and ellipticities $\bs{\ellipticity}_{\rm obs}$ and $\bs{\ellipticity}_{\rm PSF}$ of the observed image and the PSF:
\begin{eqnarray}
\bs{\ellipticity} & = & \frac{\left[F_{11}^{\rm obs}-F_{11}^{\rm PSF}-F_{22}^{\rm obs}+F_{22}^{\rm PSF}\,;\,2(F_{12}^{\rm obs}-F_{12}^{\rm PSF})\right]}{F_{11}^{\rm obs}-F_{11}^{\rm PSF}+F_{22}^{\rm obs}-F_{22}^{\rm PSF}}\nonumber\\
\label{eq:propag1}
 & = & \frac{ \bs{\ellipticity}_{\rm obs}\size^2_{\rm obs}-\bs{\ellipticity}_{\rm PSF}\size^2_{\rm PSF} }{ \size^2_{\rm obs} - \size^2_{\rm PSF} }\;.
\end{eqnarray}
Consider now that we have estimators of $\bs{\ellipticity}_{\rm PSF}$ and $\size^2_{\rm PSF}$ and the estimated values have small errors $\bs{\delta\ellipticity}_{\rm PSF}$ and $\delta\left(\ssqobj{PSF}\right)$ with respect to the true values. We can find the deviation from the truth of a single measurement of one ellipticity component $\delta\ellipticity_{{\rm gal},i}$ by differentiation:
\be
\delta\ellipticity_{{\rm gal},i} \approx
\frac{\partial\ellipticity_{{\rm gal},i}}{\partial\size^2_{\rm PSF}}\delta\size^2_{\rm PSF} +
\frac{\partial\ellipticity_{{\rm gal},i}}{\partial\ellipticity_{{\rm PSF},i}}\delta\ellipticity_{{\rm PSF},i}
\ee
which can be expanded to give 
equation \ref{eq:errorpropagationpsfintogal} and when combined with equation \ref{eq:sigmasysdef1}:
\begin{eqnarray}
\variance{ \bs{\delta\ellipticity}^{\rm sys} } & = &
       \left( \variance{ \frac{\bs{\ellipticity}_{\rm gal}}{\size^2_{\rm gal}} } + \variance{ \frac{\bs{\ellipticity}_{\rm PSF}}{\size^2_{\rm gal}} } \right)
       \variance{ \delta\left(\ssqobj{PSF}\right) } \, \nonumber\\
 & & + \left<\left(\frac{\size_{\rm PSF}}{\size_{\rm gal}}\right)^4\right>\variance{\bs{\delta\ellipticity}_{\rm PSF}}
\end{eqnarray}
which leads to 
equation \ref{eq:sigmasysdef} when assuming that the ellipticity and size of the galaxy are uncorrelated (equation \ref{eq:assumption}) and using equations \ref{eq:sigmasizedef} and \ref{eq:sigmaelldef}.

%%%%%%%%%%%%%%%%%%%%%%%%%%%%%%%%%%
%%%%%%%%%%%%%%%%%%%%%%%%%%%%%%%%%
%%%%%%%%%%%%%%%%%%%%%%%%%%%%%%%%%%

\section{Complexity factors for shapelet models}
\label{sec:appendixshapelets}

As stated in section \ref{sec:shapelets}, any object can be described by the sum of polar shapelet functions $\chi_{n,m}$ weighted by a set of complex coefficients $f_{n,m}$ (see equation \ref{eq:shap}). 
The scale parameter $\beta$ and the center of the basis can be tuned to get a sparse description of the object. Most of the time, a good choice is to take $\beta^2=\size^2/2$ and the center of the basis corresponding to the centroid of the surface brightness. Then, for simple objects like stars with reasonable substructures and tails, 
the required number of coefficients is typically less than 20. 
The `diamond' configuration, defined for $n_{\rm max}$ even, consists in setting $f_{n,m}=0$ when $\left|m\right|>n_{\rm max} - n$.

For given values of $\beta$ and the center of the basis, equations 50, 54 and 55 of \cite{2005MNRAS.363..197M} can be summarised by:
\begin{eqnarray}
\fluxtot & = & \sqrt{4\pi}\,\beta\times S_0\nonumber\\
\size^2 & = & \frac{\sqrt{16\pi}\,\beta^3}{\fluxtot}S_1\nonumber\\
\label{eq:sizedefshapelets}
 & = & 2\beta^2\times S_1 / S_0\\
\ellipticity & = & \frac{\sqrt{16\pi}\,\beta^3}{\fluxtot\,\size^2}S_2\nonumber\\
\label{eq:elldefshapelets}
 & = & S_2 / S_1
\end{eqnarray}
with:
\begin{eqnarray}
S_0 & = & \sum_{n\,\textrm{even}=0}^{n_{\textrm{max}}}f_{n0}\\
S_1 & = & \sum_{n\,\textrm{even}=0}^{n_{\textrm{max}}}(n+1)f_{n0}\\
S_2 & = & \sum_{n\,\textrm{even}=2}^{n_{\textrm{max}}}\sqrt{n(n+2)}f_{n2}\;.
\end{eqnarray}
From equations \ref{eq:sizedefshapelets}, \ref{eq:elldefshapelets} and \ref{eq:sigmap}, it follows that:
\begin{eqnarray}
\label{eq:sigsize}
\sigma[\size^2] & = & \frac{1}{\snr}\times 2\,\beta^2\times
\psi_{\size^2}
\\
\label{eq:sigell}
\sigma[\ellipticity] & = & \frac{1}{\snr}\times \frac{2\,\beta^2}{\size^2}\times
\psi_{\ellipticity}
\end{eqnarray}
with:
\begin{eqnarray}
\psi_{\size^2}
& = & \frac{1}{N/2+1}\sum_{n\,\textrm{even}=0}^{n_{\textrm{max}}}\left[(n+1-\frac{\size^2}{2\beta^2})^2\right]
\label{eq:s4}
\end{eqnarray}
and, for the non-diamond configuration:
\begin{eqnarray}
\psi_{\ellipticity}^2
& = & \frac{1}{N/2+1}\sum_{n\,\textrm{even}=0}^{n_{\textrm{max}}}\left[n(n+2)+\ellipticity^2(n+1)^2\right]\nonumber\\
\label{eq:s3}
 & = & \frac{N(N+4)}{3}+\ellipticity^2\left(\frac{N}{3}\left(N+4\right)+1\right)
\end{eqnarray}
or alternatively, if the diamond configuration is used:
\begin{eqnarray}
\psi_{\ellipticity}^2
& = & \frac{1}{N/2+1}\sum_{n\,{\rm even}=0}^{n_{\rm max}-2}\left[n(n+2)\right]+\ellipticity^2\sum_{n\,{\rm even}=0}^{n_{\rm max}}\left[(n+1)^2\right]\nonumber\\
\label{eq:s3diamond}
 & = & \frac{N(N-2)}{3}+\ellipticity^2\left(\frac{N}{3}\left(N+4\right)+1\right)
\end{eqnarray}
where $N$ is the largest even integer lower than or equal to $n_{\rm max}$. Note that $\sigma[\ellipticity]$ is the standard deviation of one component of $\bs{\ellipticity}$, according to our definition given in equation \ref{eq:sigmaelldef}.

To obtain equations \ref{eq:psidef1} to \ref{eq:psidef3}, we 
make two simplifications:
\begin{enumerate}
\item We neglect the term proportinal to $\ellipticity^2$ in
equations~\ref{eq:s3} and~\ref{eq:s3diamond}. This is justified by the `small ellipticity' assumption adopted all along this paper.
\item We choose $\size^2=2\beta^2$ in equations~\ref{eq:sigsize}, \ref{eq:sigell} and \ref{eq:s4}.
\end{enumerate}

\end{document}